\documentclass[final,5p,times,twocolumn,authoryear]{elsarticle}

\usepackage{algpseudocode}

\usepackage{algorithm}

\usepackage{tcolorbox}

\usepackage{amssymb}

\usepackage{amsmath}

\usepackage{listings}

\usepackage{hyperref}

\usepackage{booktabs}

\usepackage{makecell}

\usepackage{framed}

\usepackage{pifont}

\usepackage{xcolor}

\journal{The Journal of Systems \& Software}

\definecolor{delm}{RGB}{20,105,176}

\definecolor{numb}{RGB}{106,109,32}

\lstdefinelanguage{Json}{
  postbreak=\raisebox{0ex}[0ex][0ex]{\ensuremath{\color{gray}\hookrightarrow\space}},
  showspaces=false,
  showtabs=false,
  breakatwhitespace=true,
  breaklines=true,
  upquote=true,
  morestring=[b]", literate= *{0}{{{\color{numb}0}}}{1}
    {1}{{{\color{numb}1}}}{1}
    {2}{{{\color{numb}2}}}{1}
    {3}{{{\color{numb}3}}}{1}
    {4}{{{\color{numb}4}}}{1}
    {5}{{{\color{numb}5}}}{1}
    {6}{{{\color{numb}6}}}{1}
    {7}{{{\color{numb}7}}}{1}
    {8}{{{\color{numb}8}}}{1}
    {9}{{{\color{numb}9}}}{1}
    {\{}{{{\color{delm}{\{}}}}{1}
    {\}}{{{\color{delm}{\}}}}}{1}
    {[}{{{\color{delm}{[}}}}{1} {]}{{{\color{delm}{]}}}}{1} }

\begin{document}

  \begin{frontmatter}

    \title{OOPS: Automated generation of REST API specification via LLMs}

    \author[SCSE,JSI]{Hao Chen}

    \author[SCSE,JSI]{Yunchun Li}

    \author[SCST,JSI]{Chen Chen}

    \author[SCSE,JSI]{Fengxu Lin}

    \author[SCSE,LNT]{Wei Li \corref{cor1}}
    \ead{liw@buaa.edu.cn}

    \affiliation[SCSE]{organization={School of Computer Science and Engineering, Beihang University},city={Beijing},country={China}}

    \affiliation[SCST]{organization={School of Cyber Science and Technology, Beihang University},city={Beijing},country={China}}

    \affiliation[JSI]{organization={Sino-German Joint Software Institute, Beihang University},city={Beijing},country={China}}

    \affiliation[LNT]{organization={Beijing Key Laboratory of Networking Technology},city={Beijing},country={China}}

    \cortext[cor1]{Corresponding author}

\begin{abstract}
REST APIs, based on the REpresentational State Transfer (REST) architecture, are the primary type of Web API. The OpenAPI Specification (OAS) serves as the de facto standard for describing REST APIs and is crucial for multiple software engineering tasks. Automated OAS generation can help developers identify and correct issues in manually maintained OAS, but existing approaches rely on technology-specific rules and human expert intervention. LLMs' powerful code understanding capabilities offer the potential to overcome these limitations, but introduce additional challenges such as context length limitations and hallucinations. To address these challenges, we propose OOPS, the first technology-agnostic approach that leverages LLM-based static analysis of server code for OAS generation. Through an LLM agent workflow comprising two key steps, endpoint method extraction and OAS generation, OOPS eliminates the need for technology-specific rules or human expert intervention. By constructing an API dependency graph, it establishes necessary file associations to address LLMs' context length limitations. By multi-stage generation and self-refine, it mitigates both syntactic and semantic hallucinations during OAS generation. We evaluated OOPS on 12 real-world REST APIs spanning 5 programming languages and 8 development frameworks. Experimental results demonstrate that OOPS accurately generates high-quality OAS for REST APIs implemented with diverse technologies, achieving an average F1-score exceeding 98\% for endpoint method inference, 97\% for both request parameter and response inference, and 92\% for parameter constraint inference. The input tokens average below 5.6K with a maximum of 16.13K, while the output tokens average below 0.9K with a maximum of 7.63K. \end{abstract}


  \end{frontmatter}

\section{Introduction}\label{section-1}

REpresentational State Transfer (REST) architecture is the primary approach for designing Web APIs, and APIs based on the REST architectural style are referred to as REST APIs \citep{richardson2013restful}. The OpenAPI Specification (OAS) is a REST API description standard created by the OpenAPI Initiative, encompassing two major versions: Swagger 2.0 and OpenAPI 3.0 \citep{oas}. Driven by both academia and industry, OAS has become the de facto standard for describing REST APIs \citep{gamez2017analysis}. As a machine-readable description of REST APIs, OAS plays a crucial role in modern software engineering: it not only serves as the foundation for API documentation but also supports tasks such as automated REST API testing \citep{atlidakis2019restler,arcuri2019restful} and vulnerability detection \citep{deng2023nautilus,du2024vulnerability}, as well as integration with LLM frameworks such as LangChain and Google Agentic Development Kit (ADK) \citep{lazar2025generating}, enabling LLM agents to access real-time data and external services.

Currently, developers primarily create OAS manually or generate them through tools such as Gin-Swagger \citep{gin-swagger} and Springfox \citep{springfox} by adding technology-specific annotations or comments in code. However, research has shown that developer-maintained OAS may be outdated, inaccurate, or incomplete \citep{kim2022automated,martin2022online}. Automated OAS generation can help developers identify and correct errors in OAS \citep{huang2024generating,serbout2022expresso}, and create OAS for REST APIs that do not provide one. While multiple approaches have been proposed, they all have limitations. APICARV \citep{yandrapally2023carving} requires human experts to manually deploy REST APIs; Respector \citep{huang2024generating} and ExpressO \citep{serbout2022expresso} rely on technology-specific static analysis rules, supporting only Java and Express.js, resulting in limited generalizability that is difficult to transfer across programming languages and development frameworks.

In recent years, Large Language Models (LLMs), represented by OpenAI's ChatGPT, have demonstrated semantic understanding and content generation capabilities that provide new insights for various software engineering tasks \citep{hou2024large}. Researchers have successfully optimized tasks such as code generation \citep{dong2024self,mu2024clarifygpt}, code repair \citep{bairi2024codeplan,feng2025integrating}, generalizable software testing \citep{xia2024fuzz4all,bouzenia2025you}, and REST API testing \citep{kim2025llamaresttest} by constructing LLM workflows or agents. In OAS generation, LLMs have the potential to understand REST APIs implemented with diverse programming languages and development frameworks, overcoming existing methods' dependence on technology-specific manual rules, and improving generalizability without requiring manual deployment and execution of REST APIs.

Accordingly, we address the problem of LLM-based OAS generation: given the source code repository of a REST API implementation as input, the objective is to produce a valid OAS as output by leveraging LLMs to analyze code semantics. Despite the aforementioned potential, directly applying LLMs to this task faces two additional challenges: context length limitations and hallucinations. Recent techniques leveraging LLMs to statically analyze source code for OAS generation include the work by Chaplia et al.\ \citep{chaplia2024extracting} and LRASGen \citep{deng2025lrasgen}. To address these challenges, the former introduces a static analysis library specific to JavaScript, supporting only Node.js microservices; the latter requires human experts to write regular expressions for different server-side technologies to identify key files, supporting only limited programming languages and development frameworks, both at the cost of the generalizability that LLMs offer.

In this study, we propose OpenAI-compatible OpenAPI generator from Project Source (OOPS), a novel OAS generation method based on LLM agent workflows. Given the source code repository of a REST API implementation as input, OOPS invokes LLMs through an OpenAI-compatible API (which has been widely adopted, with mainstream models such as GPT, Gemini, and Qwen all accessible) to perform static analysis and generates an OAS as the output. By constructing an API dependency graph, OOPS accomplishes necessary file associations, thereby enabling cross-file analysis without exceeding LLM context length limitations. Through multi-stage generation and self-refine strategy, OOPS effectively mitigates LLM hallucinations during OAS generation, producing OAS that strictly adheres to the OpenAPI 3.0 specification. OOPS leverages the semantic understanding capabilities of LLMs across different server-side technologies, relying neither on technology-specific manual rules nor on human expert intervention, thus exhibiting technology-agnostic characteristics: it is generalizable across different programming languages and development frameworks. We evaluated OOPS's effectiveness on 8 open-source and 4 proprietary REST API implementations. Experimental results demonstrated that, compared to state-of-the-art methods, OOPS maintained controllable context overhead while exhibiting better generalization, and achieved higher precision and recall in inferring endpoint method, request parameter, response, and parameter constraint.

In summary, our main contributions are as follows:

\begin{itemize}

  \item We propose OOPS, an OAS generation method based on LLM agent workflows that effectively improves generalization across different programming languages and development frameworks. To the best of our knowledge, this is the first technology-agnostic OAS generation method based on LLM static analysis of REST API source code.

  \item We design a file association algorithm based on API dependency graphs to address the context length limitations of LLMs during endpoint method extraction.

  \item We design OAS generation strategies, including multi-stage generation and self-refine to mitigate syntactic and semantic hallucinations of LLMs during generation.

  \item We evaluate OOPS on 12 REST API implementations covering 5 programming languages and 8 development frameworks. Results show that compared to state-of-the-art OAS generation methods, OOPS achieves better performance and generalization.

\end{itemize}

The remainder of this paper is organized as follows: Section~\ref{section-2} introduces the background and research motivation; Section~\ref{section-3} describes our method OOPS in detail, including the overall architecture and two key steps; Section~\ref{section-4} presents the implementation details and experimental setup of OOPS; Section~\ref{section-5} provides experimental results and analysis; Section~\ref{section-6} discusses threats to validity; Section~\ref{section-7} reviews prior research related to our work; finally, we conclude the paper in Section~\ref{section-8}.

\section{Background and motivation}\label{section-2}

\subsection{REST API and OAS}\label{section-2-1}

REST is a software architectural style proposed in 2000 that enables the access and manipulation of resources based on CRUD operations \citep{fielding2000architectural}. Modern REST APIs are designed based on the REST architectural style, using the HTTP protocol as the transport layer and mapping CRUD operations to different HTTP methods: for example, the \texttt{GET} method is used to read resources, and the \texttt{POST} method is used to create new resources. In practice, a REST API implementation typically contains multiple endpoints, each with a different URL, which can execute different predefined operations through different HTTP methods based on request parameters or the request body. For terminology consistency, this paper follows the definition by Huang et al. \citep{huang2024generating}, using "endpoint method" to represent the combination of an endpoint and an HTTP method. REST APIs are typically implemented using development frameworks, such as Java-based Spring Boot and Jersey, Python-based Flask and FastAPI, as well as Node.js-based Express.js and Koa.

OAS uses structured documents (in JSON or YAML format) to describe REST APIs, specifying information such as available endpoint methods, request parameters, and response schemas for each endpoint method, enabling both humans and machines to understand the functionality of REST APIs without directly accessing their implementations. OAS 2.0 \citep{swagger-20} (also known as Swagger 2.0) and OAS 3.0 \citep{openapi-30} (also known as OpenAPI 3.0) are the two major versions of OAS. Figure~\ref{fig-2-3} and Figure~\ref{fig-2-4} show the endpoint method \texttt{POST /api/user/login} described in Swagger 2.0 and OpenAPI 3.0. Compared to Swagger 2.0, OpenAPI 3.0 renamed some elements, restructured the overall architecture to improve reusability, and further introduced and extended JSON Schema capabilities. For example, Swagger 2.0 describes the request body through items with \texttt{in} set to \texttt{body} in the \texttt{parameters} field, while OpenAPI 3.0 uses a separate \texttt{requestBody} field; additionally, the structure of the \texttt{responses} field differs between the two versions.

Currently, OpenAPI 3.0 has become the mainstream version of OAS \citep{api-trends}, and most academic research also chooses it as input \citep{atlidakis2019restler,du2024vulnerability} or output \citep{huang2024generating,serbout2022expresso}.

\begin{figure}[t]

\begin{lstlisting}[
  language=Json,
  basicstyle=\fontsize{6.4}{8}\selectfont\ttfamily,
  backgroundcolor=\color{white},
  keywordstyle=\color{blue!90!black},
  commentstyle=\color{black!50!white},
  stringstyle=\color{green!70!black},
  numberstyle=\tiny\color{gray},
  numbers=none,
  numbersep=5pt,
  stepnumber=1,
  showstringspaces=false,
  columns=flexible,
  tabsize=2
]
{
  "swagger": "2.0",
  "info": { "title": "Login API", "version": "1.0.0" },
  "paths": {
    "/api/user/login": {
      "post": {
        "parameters": [
          {
            "name": "body",
            "in": "body",
            "required": true,
            "schema": {
              "type": "object",
              "properties": {
                "username": { "type": "string" },
                "password": { "type": "string" }
              }
            }
          }
        ],
        "responses": {
          "401": { "description": "Unauthorized" },
          "200": {
            "description": "Successful",
            "schema": {
              "type": "object",
              "properties": {
                "token": { "type": "string" }
              }
            }
          }
        }
      }
    }
  }
}
\end{lstlisting}
   \vspace{-1em}
  \caption{Example of a Swagger 2.0 specification.}\label{fig-2-3}
\end{figure}

\begin{figure}[t]

\begin{lstlisting}[
  language=Json,
  basicstyle=\fontsize{6.4}{8}\selectfont\ttfamily,
  backgroundcolor=\color{white},
  keywordstyle=\color{blue!90!black},
  commentstyle=\color{black!50!white},
  stringstyle=\color{green!70!black},
  numberstyle=\tiny\color{gray},
  numbers=none,
  numbersep=5pt,
  stepnumber=1,
  showstringspaces=false,
  columns=flexible,
  tabsize=2
]
{
  "openapi": "3.0.0",
  "info": { "title": "Login API", "version": "1.0.0" },
  "paths": {
    "/api/user/login": {
      "post": {
        "requestBody": {
          "required": true,
          "content": {
            "application/json": {
              "schema": {
                "type": "object",
                "properties": {
                  "username": { "type": "string" },
                  "password": { "type": "string" }
                }
              }
            }
          }
        },
        "responses": {
          "401": { "description": "Unauthorized" },
          "200": {
            "description": "Successful",
            "content": {
              "application/json": {
                "schema": {
                  "type": "object",
                  "properties": {
                    "token": { "type": "string" }
                  }
                }
              }
            }
          }
        }
      }
    }
  }
}
\end{lstlisting}
   \vspace{-1em}
  \caption{Example of an OpenAPI 3.0 specification.}\label{fig-2-4}
\end{figure}

\subsection{Research motivation}\label{section-2-2}

\begin{figure*}[!t]
  \centering
  \includegraphics[width=0.86\textwidth]{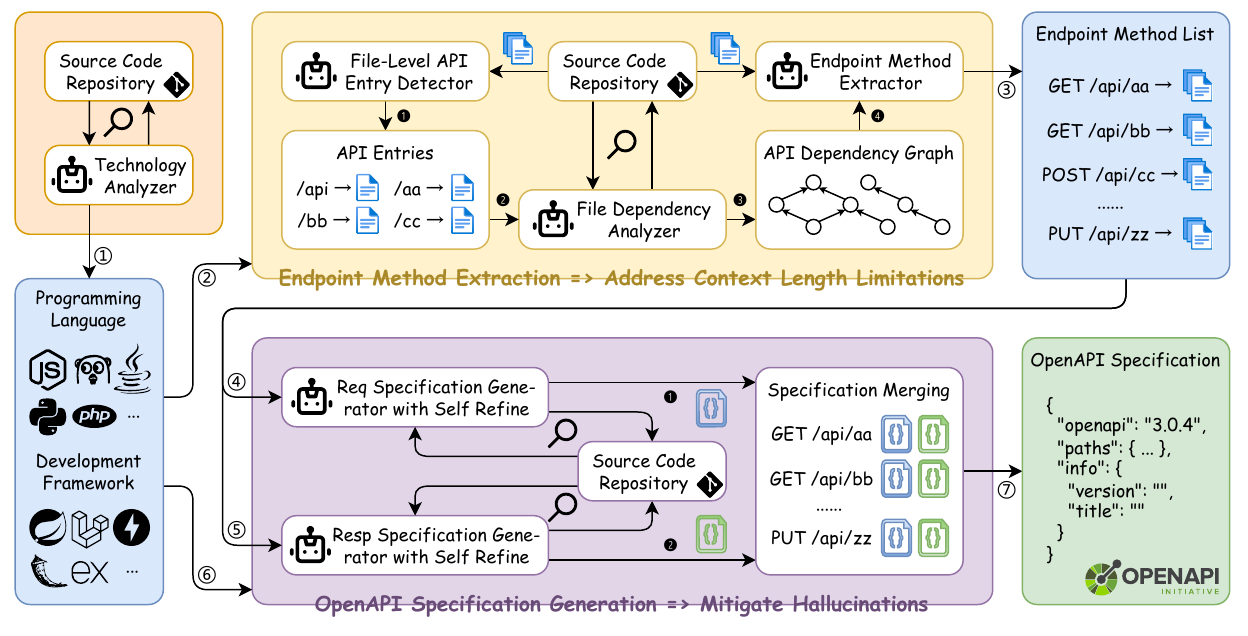}
  \vspace{-1.0em}
  \caption{Overview of the OOPS approach. The endpoint method extraction based on API dependency graph addresses context length limitations (Challenge 1). The multi-stage OpenAPI specification generation with self-refine mitigates hallucinations (Challenge 3). The approach leverages LLMs' semantic understanding capabilities to eliminate dependence on technology-specific manual rules (Challenge 2).}\label{fig-3-1}
\end{figure*}

In recent years, automated OAS generation methods based on static analysis \citep{huang2024generating,lercher2024generating} have gained widespread attention due to their ability to operate without requiring developers to modify source code or deploy REST APIs. However, these methods target specific programming languages or development frameworks, making them unable to be extended to other server-side technologies. Currently, LLMs, as a major breakthrough in generative artificial intelligence, have demonstrated strong capabilities across various downstream tasks \citep{hou2024large} such as code understanding and code generation. For the task of OAS generation, the ability of LLMs to understand code in different programming languages provides a new approach for achieving technology-agnostic solutions. For example, LLMs can understand REST API implementations based on different server-side technologies and generate the Swagger 2.0 specification shown in Figure~\ref{fig-2-3} or the OpenAPI 3.0 specification shown in Figure~\ref{fig-2-4} according to the OAS version specified in the prompt.

However, applying LLMs to OAS generation faces three challenges: context length limitations (\textbf{Challenge 1}), dependence on technology-specific manual rules (\textbf{Challenge 2}), and hallucinations (\textbf{Challenge 3}). Challenge 1 and Challenge 2 stem from the fact that API endpoint implementations typically involve multiple code files. For example, in Spring Boot, the request body is often mapped to a class defined in another file; in FastAPI, the complete URL is often split into multiple parts defined in different files and organized using \texttt{APIRouter} instances and the \texttt{include\_router} method. If all code files are provided to LLMs for analysis, the total code volume may exceed the context length limit \citep{zan2024codes,zhang2025hierarchical,liang2025can}. If manual rules are employed to select files, the approach becomes limited to specific programming languages or development frameworks, failing to fully leverage the generalization advantages of LLMs. Challenge 3 is an inherent problem of LLMs. Specifically, in OAS generation, LLMs may output invalid JSON, confuse different OAS versions, or generate OAS that does not align with the code implementation. For instance, when generating OpenAPI 3.0 specifications, LLMs may incorrectly use items with \texttt{in} set to \texttt{body} in the \texttt{parameters} field to describe the request body structure (a syntax only applicable to Swagger 2.0).

Existing LLM-based OAS generation methods fail to address all these challenges. In addressing context length limitations, Chaplia et al. \citep{chaplia2024extracting} employed the Dependency Cruiser library to analyze dependencies among JavaScript files, thus limiting their approach to Node.js-based REST APIs. They also identified hallucinations in LLM-based OAS generation. LRASGen \citep{deng2025lrasgen} addressed context length limitations through regex-based file reference parsing, such as matching \texttt{import xxx} and \texttt{from xxx import *} for Python-based REST APIs, which similarly restricts generalizability. Moreover, they provided no analysis of LLM hallucinations. In contrast, our approach performs file association analysis through an LLM agent workflow, addressing both context length limitations and dependence on technology-specific manual rules, while mitigating various types of LLM hallucinations through a multi-stage generation and self-refine strategy.

\section{Approach}\label{section-3}

\subsection{Overview}\label{section-3-1}

Our method overview is illustrated in Figure~\ref{fig-3-1}, which comprises three steps: \textbf{server-side technology analysis}, \textbf{endpoint method extraction}, and \textbf{OpenAPI specification generation}. The server-side technology analysis step is implemented using a \textbf{technology analyzer}, an LLM agent with capabilities for directory traversal, file reading, and file searching, tasked with identifying the programming language and development framework employed by the input REST API implementation. Based on this information, OOPS utilizes a role-playing mechanism to improve the response quality of LLMs in subsequent processes. Specifically, the following role description is added to the beginning of the prompt template for each LLM:

\begin{figure}[h]
  \centering

\begin{tcolorbox}[colback=lightgray,colframe=white]

  \scriptsize You are an AI assistant specialized in analyzing web projects developed using \texttt{\{language\}} and the \texttt{\{framework\}} framework.

\end{tcolorbox}

\vspace{-0.8em}
 \end{figure}

The endpoint method extraction step analyzes code files through LLMs based on the server-side technology analysis results, extracting the list of endpoint methods contained in the REST API implementation and their associated files. This step is accomplished by a workflow composed of three LLM agents: \textbf{file-level API entry detector}, \textbf{file dependency analyzer}, and \textbf{endpoint method extractor}. This step achieves necessary file associations by constructing the API dependency graph entirely through LLMs, thereby addressing the context length limitations of LLMs while avoiding dependence on technology-specific manual rules, as detailed in Section~\ref{section-3-2}.

The OpenAPI specification generation step determines the request and response specifications for each extracted endpoint method through LLMs, corrects potential hallucinations, and ultimately integrates them to produce a complete OAS. This step is accomplished by a workflow composed of two LLM agents: \textbf{req specification generator} and \textbf{resp specification generator}, as well as a \textbf{specification merging} process. This step employs a multi-stage generation and self-refine strategy to effectively mitigate syntactic and partial semantic hallucinations in LLM-generated OAS; these optimizations are general-purpose and avoid dependence on technology-specific manual rules, as detailed in Section~\ref{section-3-3}.

\subsection{Endpoint method extraction}\label{section-3-2}

\begin{algorithm}[t!]
  \caption{Endpoint method extraction}\label{algorithm-1}
  \small\textbf{Input:} REST API implementation repository (Repository) \\
  \small\textbf{Output:} REST API endpoint methods (EndpointMethods) \renewcommand{\algorithmicindent}{1em}
  \fontsize{8pt}{9pt}\selectfont
  \begin{algorithmic}[1]

    \State{APIEntries $\gets \emptyset$}
    \State{APIUnique $\gets \emptyset$}
    \For{each file $f \in$ Repository}
      \If{not isBinaryFile($f$)}
        \State{entries $\gets$ callFileLevelAPIEntryDetector($f$) \text{// agent}}
        \For{each entry $e \in$ entries}
          \State{key $\gets$ normalizePathParameters($e.path$)}
          \State{unique $\gets$ UniqueKey(key, $f$)}
          \State{entry $\gets$ APIEntry(key, $f$, $e.path$, $e.handler$, $e.tag$)}
          \If{unique $\notin$ APIUnique}
            \State{APIEntries $\gets$ APIEntries $\cup \ \{entry\}$}
            \State{APIUnique $\gets$ APIUnique $\cup \ \{unique\}$}
          \EndIf{}
        \EndFor{}
      \EndIf{}
    \EndFor{}

    \State{}

    \State{$G \gets (V, E)$ \text{// API dependency graph}}
    \State{$V \gets \emptyset$ \text{// graph nodes}}
    \State{$E \gets \emptyset$ \text{// graph edges}}
    \State{$F2E \gets$ map() \text{// file to entries mapping}}
    \For{each entry $e \in$ APIEntries}
      \State{$F2E[e.file] \gets$ getOrDefault($F2E$, $e.file$, $\emptyset$) $\cup \ \{e\}$}
      \State{$V \gets V \cup \ \{e.file\}$}
    \EndFor{}
    \For{each entry $e \in$ APIEntries}
      \If{$e.tag = REF$}
        \State{refs $\gets$ callFileDependencyAnalyzer($e$) \text{// agent}}
        \For{each file $f \in$ refs}
          \If{$f \notin V$ or $e.handler \notin$ getHandlers($F2E[f]$)}
            \State{entry $\gets$ APIEntry("", $f$, "", $e.handler$, LOCAL)} \State{$F2E[f] \gets$ getOrDefault($F2E$, $f$, $\emptyset$) $\cup \ \{entry\}$}
            \State{$V \gets V \cup \ \{f\}$}
          \EndIf{}
          \State{$E \gets E \cup \ \{(f, e.file)\}$}
        \EndFor{}
      \EndIf{}
    \EndFor{}

    \State{}

    \State{EndpointMethods $\gets \emptyset$}
    \For{each entry $e \in$ APIEntries}
      \If{$e.tag = LOCAL$}
        \State{subgraph $\gets$ makeSubgraph($G$, getDescendants($G$, $e.file$) $\cup \ e.file$)}
        \State{ordered $\gets$ topologicalSort(subgraph)}
        \State{path, methods $\gets$ callEndpointMethodExtractor(ordered) \text{// agent}}
        \For{each method $m \in$ methods}
          \State{target $\gets$ EndpointMethod(path, $m$, $e$, ordered)}
          \State{EndpointMethods $\gets$ EndpointMethods $\cup \ \{target\}$}
        \EndFor{}
      \EndIf{}
    \EndFor{}

    \State{\Return{EndpointMethods}}
  \end{algorithmic}
\end{algorithm}

As shown in Figure~\ref{fig-3-1}, OOPS invokes the \textbf{file-level API entry detector} to extract API entries from each code file (\ding{202}) as the first step in the endpoint method extraction workflow. An \textbf{API entry} is defined as a triple $APIEntry = (Path, Handler, Tag)$, where $Path$ represents a partial or complete request path, $Handler$ represents the function, class, or object that processes requests matching the path, and $Tag$ is an enumeration value that marks whether the $Handler$ implementation is located in the current file (LOCAL) or imported from other files (REF). API entries model the routing mechanism of REST API development frameworks: binding $Path$ with $Handler$ to either process matching requests or dispatch requests to other Handlers. Although routing implementations vary across programming languages and development frameworks, their core semantics can be uniformly mapped onto the technology-agnostic abstraction of API entries. Therefore, instead of requiring technology-specific parsing rules for each server-side technology, OOPS leverages the code semantic understanding capability of LLMs to directly detect API entries conforming to this unified abstraction from code files. As shown in lines 1--16 of Algorithm~\ref{algorithm-1}, OOPS normalizes path parameters using \texttt{\{\}} and deduplicates based on the normalized $Path$ and file location for each identified API entry.

As the second and third steps in the endpoint method extraction workflow, OOPS analyzes the dependencies between API entries located in different files (\ding{203}) and then constructs an \textbf{API dependency graph} to associate files (\ding{204}). The API dependency graph is defined as a directed acyclic graph $G = (V, E)$, where $V$ is the node set and $E$ is the edge set. Each node $v \in V$ represents a file; each edge $(v_1, v_2) \in E$ indicates that an API entry in file $v_2$ references a $Handler$ implemented in file $v_1$, meaning that information from file $v_2$ is required to determine the complete request path of the corresponding API entry in file $v_1$.

\begin{figure}[!t]
  \centering
  \includegraphics[width=1.00\columnwidth]{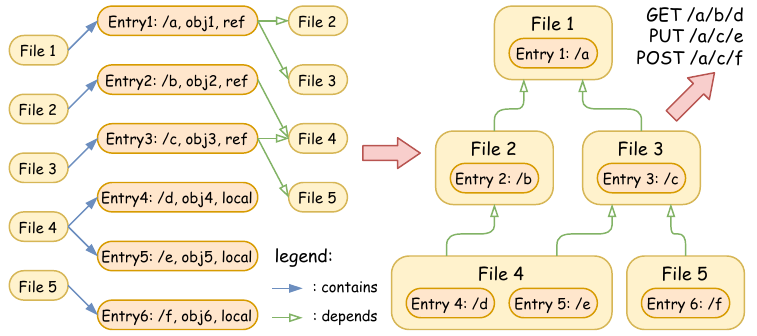}
  \vspace{-1.2em}
  \caption{An example of an API dependency graph.}\label{fig-3-2}
\end{figure}

As shown in lines 18--25 of Algorithm~\ref{algorithm-1}, OOPS creates a node for each file containing API entries. Then, as shown in lines 26--38 of Algorithm~\ref{algorithm-1}, OOPS invokes the \textbf{file dependency analyzer} to analyze the list of files where the $Handler$ of each API entry with Tag REF is implemented. For each file in the list, OOPS adds an edge from that file to the file containing the current API entry. For example, in Figure~\ref{fig-3-2}, requests with prefix \texttt{/a} are processed by Entry 1 in File 1 and dispatched to specific implementations in File 2 or File 3 based on the path; thus, edges from File 2 to File 1 and from File 3 to File 1 exist in the API dependency graph. In particular, if a file in the list is not yet present in the API dependency graph, OOPS creates a new node for that file and adds an API entry located in that file with an empty $Path$, the $Handler$ inherited from the corresponding API entry with Tag REF, and the Tag set to LOCAL.

As the last step in the endpoint method extraction workflow, OOPS integrates path fragments from different files based on the API dependency graph to obtain complete request paths and HTTP methods, thereby generating the endpoint method list (\ding{205}). As shown in lines 40--51 of Algorithm~\ref{algorithm-1}, OOPS traverses API entries with Tag LOCAL. For each entry, OOPS constructs a subgraph of the API dependency graph based on its node and all descendant nodes, and inputs each node to the \textbf{endpoint method extractor} in topological order to obtain the merged complete request path and a list of HTTP methods. Each combination of the request path and an HTTP method in the list constitutes an endpoint method. For example, in Figure~\ref{fig-3-2}, the descendant nodes of File 5, where Entry 6 is located, include File 3 and File 1. After topologically sorting the subgraph formed by these three nodes, the files input to the endpoint method extractor in sequence are File 5, File 3, and File 1. Based on these files, the endpoint method extractor can determine that the complete request path corresponding to Entry 6 is \texttt{/a/c/f}, with the HTTP method \texttt{POST}. The API dependency graph explicitly models cross-file routing dependencies, enabling OOPS to provide the LLM with only the files necessary to determine each endpoint method, rather than the entire repository, effectively addressing the challenge of context length limitations.

\subsubsection*{File-level API entry detector}

The file-level API entry detector is an LLM agent tasked with detecting API entries in the analyzed file. Given that LLMs impose token limits on their outputs, a single inference may not cover all API entries. Therefore, the agent operates in an iterative loop. In each iteration, OOPS appends the identified API entries to the prompt, guiding the LLM to focus on previously overlooked parts until no new API entries are discovered. We employ chain-of-thought techniques to enhance the LLM's reasoning quality and use tool calling to constrain its outputs to structured content conforming to a predefined JSON Schema. The format of the prompt template is as follows:

\begin{figure}[h]
  \centering

\begin{tcolorbox}[colback=lightgray,colframe=white]

  \small\textbf{File-Level API Entry Detection Prompt (Simplified)}
  \vspace{0.6em}
  \hrule

  \vspace{-0.5em}
  \begin{center}
  \small\textbf{Task Description}
  \end{center}
  \vspace{-0.8em}

  \scriptsize You need to extract API entries from file \texttt{\{file\}}, its content is \texttt{\{content\}}. The definition of an API entry is \ldots

  \vspace{-1.0em}
  \begin{center}
  \small\textbf{Self Refine}
  \end{center}
  \vspace{-1.0em}

  \scriptsize These are the API entries you've found \texttt{\{entries\}}. Continue searching for others. Output an empty list if no additional \ldots

  \vspace{-1.0em}
  \begin{center}
  \small\textbf{CoT \& Structured Output}
  \end{center}
  \vspace{-1.0em}

  \scriptsize Think step by step, and use the output tool to provide your thought process, along with a list containing \ldots

\end{tcolorbox}

\vspace{-0.8em}
 \end{figure}

Furthermore, we observe that certain files can be determined not to contain API entries solely based on their file paths without reading their contents, such as configuration files or unit test files. Therefore, OOPS performs pre-filtering on non-binary files through an additional LLM call. Only when the judgment indicates the potential presence of API entries does OOPS conduct an in-depth analysis of the file content. The format of the prompt template is as follows:

\begin{figure}[h]
  \centering

\begin{tcolorbox}[colback=lightgray,colframe=white]

  \small\textbf{File Filtering Prompt (Simplified)}
  \vspace{0.6em}
  \hrule

  \vspace{-0.5em}
  \begin{center}
  \small\textbf{Task Description}
  \end{center}
  \vspace{-0.8em}

  \scriptsize You need to analyze whether file \texttt{\{file\}} might contain definitions of REST API endpoints, rather than \ldots

  \vspace{-1.0em}
  \begin{center}
  \small\textbf{Structured Output}
  \end{center}
  \vspace{-1.0em}

  \scriptsize Use the output tool to output your result ("true" or "false"), along with the reasoning behind your judgment.

\end{tcolorbox}

\vspace{-0.8em}
 \end{figure}

During the API entry detection step, the primary hallucination exhibited by LLMs is inconsistent path parameter formats, i.e., failing to use the unified \texttt{/\{param\}} format as required but instead outputting variant forms such as \texttt{/:param} or \texttt{/<param>} influenced by different implementation approaches. To address this, OOPS performs normalization on path parameters, uniformly converting them to the \texttt{/\{param\}} format.

\subsubsection*{File dependency analyzer}

The file dependency analyzer is an LLM agent tasked with locating referenced files. We provide the agent with tool functions, including listing directories, reading files, and searching files. We observe that the primary hallucination of this agent lies in incorrect file paths, i.e., LLMs failing to strictly follow the prompt requirements to output canonical paths relative to the project root directory. To address this, we design a self-refine strategy: when invalid file paths are detected in the LLM's outputs, OOPS provides specific error feedback and requests corrections until all file paths are valid. The format of the prompt template is as follows:

\begin{figure}[h]
  \centering

\begin{tcolorbox}[colback=lightgray,colframe=white]

  \small\textbf{File Dependency Analysis Prompt (Simplified)}
  \vspace{0.6em}
  \hrule

  \vspace{-0.5em}
  \begin{center}
  \small\textbf{Task Description}
  \end{center}
  \vspace{-0.8em}

  \scriptsize In file \texttt{\{file\}}, I found that `\texttt{\{handler\}}` is imported to process HTTP requests to path `\texttt{\{path\}}`. Your task is to find which files \ldots

  \vspace{-1.0em}
  \begin{center}
  \small\textbf{Self Refine}
  \end{center}
  \vspace{-1.0em}

  \scriptsize These are the file paths you've found \texttt{\{files\}}. Some of them are not relative to the project root directory \texttt{\{errors\}}. Try to fix \ldots

  \vspace{-1.0em}
  \begin{center}
  \small\textbf{Structured Output}
  \end{center}
  \vspace{-1.0em}

  \scriptsize Use the output tool to provide your reasoning and a list of the files you found, with each item presented as a file path \ldots

\end{tcolorbox}

\vspace{-2.0em}
 \end{figure}

\subsubsection*{Endpoint method extractor}

The endpoint method extractor is an LLM agent tasked with integrating path fragments and determining HTTP methods. We similarly employ chain-of-thought techniques and structured output control. Considering the fixed set of HTTP methods, to reduce omissions in the LLM's outputs, OOPS requires the LLM to output whether each HTTP method is acceptable, rather than outputting a list of acceptable HTTP methods. Additionally, given that most REST API development frameworks allow developers not to explicitly specify HTTP methods (e.g., \texttt{@RequestMapping} in Spring Boot and \texttt{app.route} in Flask), OOPS sets an additional \texttt{ALL} method, allowing the LLM to avoid individually marking each HTTP method. This agent also suffers from the hallucination of inconsistent path parameter formats, for which OOPS applies the same normalization strategy. The format of the prompt template is as follows:

\begin{figure}[h]
  \centering

\begin{tcolorbox}[colback=lightgray,colframe=white]

  \small\textbf{Endpoint Method Extraction Prompt (Simplified)}
  \vspace{0.6em}
  \hrule

  \vspace{-0.5em}
  \begin{center}
  \small\textbf{Task Description}
  \end{center}
  \vspace{-0.8em}

  \scriptsize In file \texttt{\{file\}}, I found that `\texttt{\{handler\}}` is imported to process HTTP requests to path `\texttt{\{path\}}`, here are the relevant files I found: \texttt{\{dependencies\}}. You need to determine the full path and all acceptable HTTP methods \ldots

  \vspace{-1.0em}
  \begin{center}
  \small\textbf{CoT \& Structured Output}
  \end{center}
  \vspace{-1.0em}

  \scriptsize Think step by step, and use the output tool to provide your thought process, along with an object containing \ldots

\end{tcolorbox}

\vspace{-2.0em}
 \end{figure}

\subsection{OpenAPI Specification generation}\label{section-3-3}

As shown in Figure~\ref{fig-3-1}, to mitigate hallucinations in LLM-generated OAS, OOPS adopts a \textbf{multi-stage generation strategy}: through a workflow consisting of two LLM agents, namely the \textbf{req specification generator} and the \textbf{resp specification generator}, the complex OAS generation task is broken down into simpler subtasks that separately generate the request specification (\ding{202}) and response specification (\ding{203}) for each individual endpoint method. Subsequently, OOPS performs \textbf{specification merging}. As shown in lines 27--36 of Algorithm~\ref{algorithm-2}, OOPS integrates the request and response specifications of each endpoint method following the two-level indexing structure of OAS from Paths objects to Path Item objects to Operation objects, thereby producing a complete OAS that conforms to the Swagger 2.0 specification. Finally, OOPS leverages Swagger Codegen \citep{swagger-codegen} to convert it into the more widely adopted OpenAPI 3.0.4 version.

During the generation of request and response specifications, OOPS performs inline expansion of all references, which leads to redundant schema definitions in the final OAS. To optimize the structure of the OAS, OOPS implements reference reconstruction during specification merging. As illustrated in the rebuildReference procedure and lines 38--44 of Algorithm~\ref{algorithm-2}, OOPS traverses each Schema object in request parameters and responses, computes a hash value of its serialized representation as a unique identifier, migrates duplicate Schema objects to \texttt{components/schemas}, and replaces the original Schema with a reference pointing to \texttt{\#/components/schemas/\{hash\}}.

\begin{table}[t]
  \scriptsize
  \centering
  \caption{Common JSON syntax errors observed in LLM-generated outputs.}\label{table-3-1}
  \vspace{0.2em}
  \begin{tabular}{@{} l l l @{}}
    \toprule
    \textbf{Type}
    & \textbf{Description}
    & \textbf{Example}
    \\
    \midrule
    Backtick   & Output is enclosed in backticks.  & \texttt{\textasciigrave\textasciigrave\textasciigrave\{"key":"value"\}\textasciigrave\textasciigrave\textasciigrave} \\ Comment    & JSON string contains comments.    & \texttt{\{"key":"value"\} // text} \\ List Comma & Comma at the end of the list.     & \texttt{[1,2,3,4,5,6,7,]} \\ Dict Comma & Comma at the end of the dict.     & \texttt{\{"key1":1,"key2":2,\}} \\ Quote      & Incorrect use of quotation marks. & \texttt{\{“key”:'value'\}} \\ Reference  & Referenced a non-existent object. & \texttt{\{"key":\{"\$ref":"\#/nil"\}\}} \\ \bottomrule
  \end{tabular}
\end{table}

\begin{figure}[t]
  \centering
  \begin{minipage}[b]{0.36\columnwidth}

\begin{lstlisting}[
  language=Json,
  basicstyle=\fontsize{6.4}{8}\selectfont\ttfamily,
  backgroundcolor=\color{white},
  keywordstyle=\color{blue!90!black},
  commentstyle=\color{black!50!white},
  stringstyle=\color{red!70!black},
  numberstyle=\tiny\color{gray},
  numbers=none,
  numbersep=5pt,
  stepnumber=1,
  showstringspaces=false,
  columns=flexible,
  tabsize=2
]
[
  {
    "name": "username",
    "in": "body",
    "required": true,
    "type": "string"
  },
  {
    "name": "password",
    "in": "body",
    "required": true,
    "type": "string"
  }
]
\end{lstlisting}
 \end{minipage}
  \begin{minipage}[b]{0.52\columnwidth}

\begin{lstlisting}[
  language=Json,
  basicstyle=\fontsize{6.4}{8}\selectfont\ttfamily,
  backgroundcolor=\color{white},
  keywordstyle=\color{blue!90!black},
  commentstyle=\color{black!50!white},
  stringstyle=\color{green!70!black},
  numberstyle=\tiny\color{gray},
  numbers=none,
  numbersep=5pt,
  stepnumber=1,
  showstringspaces=false,
  columns=flexible,
  tabsize=2
]
[
  {
    "name": "body",
    "in": "body",
    "required": true,
    "schema": {
      "type": "object",
      "properties": {
        "username": { "type": "string" },
        "password": { "type": "string" }
      }
    }
  }
]
\end{lstlisting}
 \end{minipage}
  \vspace{-1em}
  \caption{Common OAS violations observed in LLM-generated outputs, with incorrect output on the left and correct output on the right.}\label{fig-3-3}
\end{figure}

\subsubsection*{Req/Resp specification generator}

The req specification generator and resp specification generator are two LLM agents tasked with generating specifications for requests and responses of a given endpoint method. Similar to the aforementioned agents, these two agents are also equipped with tool functions such as listing directories, reading files, and searching files, enabling LLMs to identify request parameters or response structures defined in other files. Unlike the previous agents, these two agents do not employ structured output based on tool calling, nor do they utilize the native structured output capabilities of LLMs. This is because the OAS Schema contains numerous optional fields, recursive definitions, and deeply nested structures, making it far more complex than the output schemas described earlier. Existing LLMs struggle to effectively handle such complex structures \citep{structured-outputs}, often resulting in optional fields being ignored or outputs conflicting with the Schema definition. Therefore, these two agents do not rely on structured output mechanisms, but instead leverage the OAS semantic knowledge acquired by LLMs during pretraining to directly output JSON strings, which are then parsed and corrected. The format of the prompt template is as follows, differing only in the generation target:

\begin{figure}[h]
  \centering

\begin{tcolorbox}[colback=lightgray,colframe=white]

  \small\textbf{Specification Generation Prompt (Simplified)}
  \vspace{0.6em}
  \hrule

  \vspace{-0.5em}
  \begin{center}
  \small\textbf{Task Description}
  \end{center}
  \vspace{-0.8em}

  \scriptsize I'm focusing on path `\texttt{\{path\}}` and method \texttt{\{method\}}, here are the relevant files I found: \texttt{\{dependencies\}}. Your task is to generate \ldots

  \vspace{-1.0em}
  \begin{center}
  \small\textbf{Self Refine}
  \end{center}
  \vspace{-1.0em}

  \scriptsize This is the specification you generated, its error is \texttt{\{error\}}, try to find the missing reference content and correct it: \texttt{\{specification\}}.

  \vspace{-1.0em}
  \begin{center}
  \small\textbf{Structured Output}
  \end{center}
  \vspace{-1.0em}

  \scriptsize You need to output in Swagger 2.0 JSON format, do not include fields that are not mentioned above, do not add any additional explanations, \ldots

\end{tcolorbox}

\vspace{-0.8em}
 \end{figure}

\begin{algorithm}[t!]
  \caption{OpenAPI specification generation}\label{algorithm-2}
  \small\textbf{Input:} REST API endpoint methods (EndpointMethods) \\
  \small\textbf{Output:} OpenAPI 3.0.4 specification (OpenAPISpec) \renewcommand{\algorithmicindent}{1em}
  \fontsize{8pt}{9pt}\selectfont
  \begin{algorithmic}[1]

    \Procedure{generateSpec}{agentFunction}
      \State{specs $\gets$ map()}
      \For{each endpoint method $em \in$ EndpointMethods}
        \State{err $\gets$ null}
        \While{true}
          \State{spec $\gets$ fixSyntaxError(agentFunction($em$, err))}
          \State{spec, err $\gets$ checkAndExpandReference(spec)}
          \If{err = null}
            \State{\textbf{break}}
          \EndIf{}
        \EndWhile{}
        \State{specs[$(em.path, em.method)$] $\gets$ convertToVersion20(spec)}
      \EndFor{}
      \State{\Return{specs}}
    \EndProcedure{}

    \State{}

    \Procedure{rebuildReference}{objectList, schemaPool}
      \For{each object $i \in$ objectList}
        \If{hasAttribute($i$, "schema")} \State{key $\gets$ md5($i.schema$)}
          \State{schemaPool[key] $\gets i.schema$}
          \State{$i.schema \gets$ makeReference("\#/components/schemas/" + key)} \EndIf{}
      \EndFor{}
    \EndProcedure{}

    \State{}

    \State{ReqSpecs $\gets$ generateSpec(callReqSpecificationGenerator) \text{// agent}}
    \State{RespSpecs $\gets$ generateSpec(callRespSpecificationGenerator) \text{// agent}}
    \State{Draft $\gets$ EmptyOpenAPISpecVersion20()}
    \For{each endpoint method $em \in$ EndpointMethods}
      \State{Draft.paths[$em.path$] $\gets$ getOrDefault(Draft.paths, $em.path$, map())}
      \State{operation $\gets$ EmptyOperationObject()}
      \State{operation.parameters $\gets$ ReqSpecs[$(em.path, em.method)$]}
      \State{operation.responses $\gets$ RespSpecs[$(em.path, em.method)$]}
      \State{Draft.paths[$em.path$][$em.method$] $\gets$ operation}
    \EndFor{}

    \State{}

    \State{OpenAPISpec $\gets$ upgradeToVersion304(Draft)}
    \State{SchemaPool $\gets$ map()}
    \State{rebuildReference(searchParameterObj(OpenAPISpec), SchemaPool)}
    \State{rebuildReference(searchMediaTypeObj(OpenAPISpec), SchemaPool)}
    \For{each key $k$, schema $s \in$ SchemaPool}
      \State{OpenAPISpec.components.schemas[$k$] $\gets s$}
    \EndFor{}

    \State{\Return{OpenAPISpec}}
  \end{algorithmic}
\end{algorithm}

During the generation of request and response specifications, the hallucination problem of LLMs manifests at both syntactic and semantic levels. Syntactic hallucinations refer to cases where LLMs fail to output parsable JSON text according to JSON syntax, with typical error types shown in Table~\ref{table-3-1}. Semantic hallucinations refer to cases where LLMs output syntactically correct JSON whose content either does not conform to the REST API semantics encoded by OAS, or reflects the LLM's misunderstanding of the source code's business logic. The former manifests as OAS violations: regardless of whether the LLM is instructed to generate Swagger 2.0 or OpenAPI 3.0 specifications, it may mix features of both versions or generate JSON structures that violate OAS constraints. For example, the incorrect output on the left side of Figure~\ref{fig-3-3} is entirely valid at the JSON level, but the LLM declares two fields as separate body parameters, violating the Swagger 2.0 constraint that only one body parameter is allowed. This essentially reflects the LLM's confusion over the semantic differences among different parameter locations (\texttt{in}). The latter manifests as factual errors inconsistent with actual source code behavior, such as incorrect parameter locations, incorrect response status codes, and missing or fabricated parameters for endpoint methods.

To mitigate syntactic hallucinations, OOPS implements a \textbf{self-refine} strategy. As shown in lines 4--11 of Algorithm~\ref{algorithm-2}, OOPS first corrects syntax errors unrelated to references based on technology-agnostic rules (such as removing extra commas and comments), then attempts to parse and expand references. When expansion fails due to non-existent reference targets, OOPS feeds the error information back to the LLM and requires it to call tools to determine the target definition of the reference, until all references are correctly resolved, thereby obtaining syntactically correct JSON output.

Regarding semantic hallucinations, OOPS focuses on mitigating OAS violations. Through the analysis of LLM outputs, we found that the generated results are generally closer to the Swagger 2.0 specification, particularly when the prompt explicitly prohibits outputting \texttt{requestBody}. Therefore, OOPS instructs the LLM in the prompt to output a Swagger 2.0 Parameter object list or Responses object, and performs corrections on this basis to fix common errors such as those shown in Figure~\ref{fig-3-3}, ultimately producing request and response specifications that conform to the Swagger 2.0 specification.

\section{Experimental methodology}\label{section-4}

\subsection{Research questions}\label{section-4-1}

To evaluate the effectiveness of OOPS, we designed experiments to answer the following research questions:

\begin{itemize}

  \item \textbf{RQ1 (Method effectiveness):} Can OOPS effectively generate high-quality OAS? In this RQ, we compare the OAS generated by OOPS with 3 state-of-the-art methods across four dimensions: endpoint method, request parameter, response, and parameter constraint, in terms of precision, recall, and F1-score, to validate the effectiveness of OOPS in generating OAS and evaluate its applicability to different programming languages and development frameworks.

  \item \textbf{RQ2 (Ablation study):} What are the contributions of different components to the effectiveness of OOPS? In this RQ, we evaluate two variants of OOPS: (1) replacing the endpoint method extraction phase (Section~\ref{section-3-2}) with a simple ReAct-style agent, and (2) replacing the OpenAPI specification generation phase (Section~\ref{section-3-3}) with a single-pass generation approach. We compare the performance of OOPS with these two variants to analyze the contributions of each component to effectiveness.

  \item \textbf{RQ3 (Impact of different LLM choices):} How does the choice of base LLM affect the effectiveness of OOPS? In this RQ, we compare the quality of OAS generated by OOPS using 5 different base LLMs to evaluate the sensitivity of OOPS to base LLM selection.

  \item \textbf{RQ4 (LLM Context Efficiency):} Does OOPS maintain reasonable context overhead when analyzing large code repositories? In this RQ, we systematically measure the number of LLM calls, context usage, output token overhead, and the number of associated files during the endpoint method extraction and OpenAPI specification generation processes of OOPS, to evaluate how effectively OOPS addresses the context length limitations of LLMs while maintaining cost-efficiency.

\end{itemize}

\subsection{Datasets}\label{section-4-2}

\begin{table*}[t]
  \scriptsize
  \centering
  \caption{Real-world REST APIs used in the evaluation. LoC denotes lines of code, and Frontend indicates whether a web-based client is available.}\label{table-4-1}
  \vspace{0.2em}
  \begin{tabular}{@{} c c c c c c c c c c @{}}
    \toprule
    \textbf{REST API} & \textbf{Commit Hash} & \textbf{Language} & \textbf{Framework} & \textbf{Code Files} & \textbf{Total LoC} & \textbf{Average LoC} & \textbf{Frontend} & \textbf{Open Source} & \textbf{Baseline Tool(s)} \\
    \midrule
    RealWorld\footnotemark[1]    & 32a16a6 & Go & Gin & 126 & 14.0K & 111.0 & $\checkmark$ & $\checkmark$ & APICARV \\
    Lin-CMS\footnotemark[2]      & 5f56d53 & Python & Flask & 272 & 37.4K & 137.4 & $\checkmark$ & $\checkmark$ & APICARV \\
    Management                   & de10901 & Python & FastAPI & 252 & 17.2K & 68.3 & $\checkmark$ & \ding{55} & APICARV \\
    CMS                          & c760724 & JavaScript & Koa & 116 & 10.0K & 86.0 & $\checkmark$ & \ding{55} & APICARV \\
    Accountill\footnotemark[3]   & 399f2a3 & JavaScript & Express.js & 88 & 5.4K & 61.8 & $\checkmark$ & $\checkmark$ & ExpressO, APICARV \\
    ProShop\footnotemark[4]      & e23e3a8 & JavaScript & Express.js & 70 & 5.4K & 77.0 & $\checkmark$ & $\checkmark$ & ExpressO, APICARV \\
    Proxy-Print\footnotemark[5]  & 87b4d13 & Java & Spring Boot & 90 & 17.3K & 192.3 & \ding{55} & $\checkmark$ & Respector \\
    Ur-Codebin\footnotemark[6]   & 83e3b4d & Java & Spring Boot & 46 & 2.2K & 47.8 & \ding{55} & $\checkmark$ & Respector \\
    Cassandra\footnotemark[7]    & 44495f1 & Java & Jersey & 140 & 18.2K & 130.0 & \ding{55} & $\checkmark$ & Respector \\
    Education-1                  & 69b4577 & Java & Spring Boot & 399 & 33.6K & 84.3 & \ding{55} & \ding{55} & Respector \\
    Education-2                  & 47c8b41 & Go & Gin & 55 & 8.5K & 154.0 & \ding{55} & \ding{55} & No available baseline \\
    Laravel-Blog\footnotemark[8] & c88e307 & PHP & Laravel & 280 & 8.0K & 28.6 & \ding{55} & $\checkmark$ & No available baseline \\
    \bottomrule
  \end{tabular}
\end{table*}

\footnotetext[1]{RealWorld: \href{https://github.com/gothinkster/realworld}{https://github.com/gothinkster/realworld}} \footnotetext[2]{Lin-CMS: \href{https://github.com/talelin/lin-cms-flask}{https://github.com/talelin/lin-cms-flask}} \footnotetext[3]{Accountill: \href{https://github.com/panshak/accountill}{https://github.com/panshak/accountill}} \footnotetext[4]{ProShop: \href{https://github.com/bradtraversy/proshop-v2}{https://github.com/bradtraversy/proshop-v2}} \footnotetext[5]{Proxy-Print: \href{https://github.com/proxyprint/proxyprint-kitchen}{https://github.com/proxyprint/proxyprint-kitchen}} \footnotetext[6]{Ur-Codebin: \href{https://github.com/mathew-estafanous/ur-codebin-api}{https://github.com/mathew-estafanous/ur-codebin-api}} \footnotetext[7]{Cassandra: \href{https://github.com/k8ssandra/management-api-for-apache-cassandra}{https://github.com/k8ssandra/management-api-for-\ldots}} \footnotetext[8]{Laravel-Blog: \href{https://github.com/guillaumebriday/laravel-blog}{https://github.com/guillaumebriday/laravel-blog}}

To demonstrate the technology-agnostic nature of OOPS, we evaluated it on 12 REST API implementations involving 5 programming languages and 8 development frameworks. Considering that open-source REST API implementations may appear in the training data of LLMs, which could compromise the validity of our experimental results, our dataset consists of 8 open-source REST API implementations and 4 proprietary REST API implementations. Specifically, following Huang et al. \citep{huang2024generating}, we selected Proxy-Print, Ur-Codebin, and Cassandra; additionally, we searched GitHub for projects adopting different REST API development frameworks and selected RealWorld, ProShop, Accountill, Lin-CMS, and Laravel-Blog based on criteria including the number of stars, available web clients, and deployability. The other 4 proprietary REST API implementations were sourced from web applications deployed in our campus network.

We analyzed the code size of these REST API implementations using the cloc tool, excluding non-core code files such as CSS, SVG, and package-lock.json. The detailed server-side technology, code statistics, and commit versions used in our experiments are shown in Table~\ref{table-4-1}. It should be noted that some REST API implementations lack available web clients. Cassandra and Education-1 do not provide web clients. Although Ur-Codebin and Laravel-Blog provide web clients, their clients employ communication methods independent of REST APIs, making it impossible to infer OAS through crawler-based approaches. The REST APIs exposed by Education-2 are partially invoked by mobile applications and partially by other web services. For Proxy-Print, we encountered the same deployment issues as Huang et al. \citep{huang2024generating} and were unable to successfully deploy the service.

To obtain the ground truth, we engaged 2 experts with extensive experience in REST API development and OAS maintenance to independently analyze the source code of each project and, where available, reference the developer-provided OAS to manually construct the OAS for these REST API implementations. Upon completion of their analyses, the 2 experts reconciled their findings to reach consensus, ultimately establishing the OAS as the ground truth.

\subsection{Baselines}\label{section-4-3}

We selected APICARV \citep{yandrapally2023carving}, ExpressO \citep{serbout2022expresso}, and Respector \citep{huang2024generating} as baseline methods. These methods are all open-source and represent two mainstream technical approaches to OAS generation: based on web clients and based on server code. APICARV is a representative method based on web clients that requires human experts to deploy both the REST API server and its web client, write approximately 10 specific rules for data preparation and static resource filtering, and then trigger REST API calls through a crawler to infer the OAS.

ExpressO and Respector are representative methods based on server code. Unlike OOPS, they rely on traditional static analysis techniques, whereas OOPS leverages LLMs for code understanding. Specifically, ExpressO targets Express.js, contains approximately 30 manually crafted analysis rules, and requires human experts to start the REST API server; Respector targets Java-based applications, employing approximately 90 manual rules to support Spring Boot and approximately 70 rules to support Jersey, without requiring runtime human intervention. In contrast, OOPS leverages LLMs' understanding of code semantics and REST API definitions, requiring no technology-specific manual rules and no need to deploy or start services.

These baseline methods have different applicable scopes, as shown in Table~\ref{table-4-1}. We ran the open-source code of these baseline methods with default configurations to reproduce their results. For APICARV, we used its built-in Crawljax tool to perform 30 minutes of web crawling. Since all selected REST APIs have user registration and login functionalities, and APICARV's crawler cannot automatically complete these operations, we manually performed the registration and login operations after 10 minutes of crawling, followed by an additional 20 minutes of automated crawling. It should be noted that since existing LLM-based OAS generation methods have not yet released runnable open-source implementations, we were unable to conduct comparative experiments with these methods.

\subsection{Performance metrics}\label{section-4-4}

To evaluate the effectiveness of OAS generation methods, we assess the quality of generated OAS across four dimensions: endpoint method, request parameter, response, and parameter constraint, following the work of Huang et al. \citep{huang2024generating} and Deng et al. \citep{deng2025lrasgen}. Given the ambiguity in prior research regarding the definition of "correct inference", we explicitly define and refine the criteria for each dimension:

\begin{itemize}

  \item \textbf{Endpoint Method:} An endpoint method is considered correctly inferred if and only if the generated result has the same HTTP method as the ground truth and the same path after normalizing path parameters.

  \item \textbf{Request Parameter:} A request parameter is considered correctly inferred if and only if the generated result has the same parameter name, data type, and location in the request as the ground truth.

  \item \textbf{Response:} A response is considered correctly inferred if and only if the generated result matches the ground truth in status code, response type, and MIME type. Consistent with our baselines, we focus on successful responses (2XX), which describe the intended API behavior and the core data models that clients depend on.

  \item \textbf{Parameter Constraint:} We examine a total of 13 constraint types: required, pattern, maxProperties, minProperties, exclusiveMaximum, exclusiveMinimum, maximum, minimum, maxLength, minLength, maxItems, minItems, and uniqueItems. A parameter constraint is considered correctly inferred if and only if both the constraint type and constraint value in the generated result exactly match those in the ground truth. It should be noted that we automatically attach a required constraint with value true to all path parameters.

\end{itemize}

Based on the above criteria, we employ three metrics: Precision, Recall, and F1-score to quantitatively evaluate the performance of OAS generation methods. These metrics are calculated from the statistics of True Positives (TP), False Positives (FP), and False Negatives (FN), where TP represents the number of correctly inferred entities, FP represents the number of incorrectly inferred entities that do not actually exist, and FN represents the number of actually existing entities that failed to be inferred. Specifically:

\begin{itemize}

  \item \textbf{Precision:} $Precision = \frac{TP}{TP + FP}$, representing the proportion of TP among all inferred entities.

  \item \textbf{Recall:} $Recall = \frac{TP}{TP + FN}$, representing the proportion of TP among all truly existing entities.

  \item \textbf{F1-score:} $F1\text{-}score = 2 \times \frac{Precision \times Recall}{Precision + Recall}$, which is the harmonic mean of Precision and Recall, providing a comprehensive evaluation of the method's overall performance.

\end{itemize}

By calculating these three metrics across the four dimensions, we can comprehensively evaluate the performance of OAS generation methods in different aspects.

\subsection{Implementation details}\label{section-4-5}

\begin{table}[t]
  \scriptsize
  \centering
  \caption{LLMs used in the evaluation. Ctx Win denotes the context window size, and Max Out indicates the maximum number of output tokens.}\label{table-4-2}
  \vspace{0.2em}
  \begin{tabular}{@{} c @{\hskip 8pt} c @{\hskip 8pt} c @{\hskip 8pt} c @{}}
    \toprule
    \textbf{Model}
    & \textbf{Ctx Win}
    & \textbf{Max Out}
    & \textbf{Price}
    \\
    \midrule
    Qwen3-Coder-Plus \citep{qwen3-coder-plus}         & 1M   & 64K  & In \$1, Out \$5 \\
    GPT-5 mini \citep{gpt-5-mini}                     & 400K & 128K & In \$0.25, Out \$2 \\
    GPT-5 nano \citep{gpt-5-nano}                     & 400K & 128K & In \$0.05, Out \$0.4 \\
    Gemini 2.0 Flash-Lite \citep{gemini-2-flash-lite} & 1M   & 8K   & In \$0.075, Out \$0.3 \\
    Gemini 3 Flash \citep{gemini-3-flash}             & 1M   & 64K  & In \$0.5, Out \$3 \\
    \bottomrule
  \end{tabular}
\end{table}

\begin{table*}[ht]
  \scriptsize
  \centering
  \caption{Comparison of OOPS with 3 state-of-the-art OAS generation techniques on 12 REST APIs. The BL columns represent the baseline results, where different baselines are used for different REST APIs. Each REST API is annotated as REST API / Baseline, where Ac represents APICARV, Ex represents ExpressO, and Rs represents Respector. \ding{55} indicates no applicable baseline for the corresponding REST API implementation. For ProShop and Accountill, we use ExpressO as the baseline due to its better performance.}\label{table-5-1} \vspace{0.2em}
  \begin{tabular}{@{} l @{\hskip 4pt}
    c @{\hskip 4pt} c @{\hskip 8pt} c @{\hskip 4pt} c @{\hskip 8pt} c @{\hskip 4pt} c @{} p{16pt} @{}
    c @{\hskip 4pt} c @{\hskip 8pt} c @{\hskip 4pt} c @{\hskip 8pt} c @{\hskip 4pt} c @{} p{16pt} @{}
    c @{\hskip 4pt} c @{\hskip 8pt} c @{\hskip 4pt} c @{\hskip 8pt} c @{\hskip 4pt} c @{} p{16pt} @{}
    c @{\hskip 4pt} c @{\hskip 8pt} c @{\hskip 4pt} c @{\hskip 8pt} c @{\hskip 4pt} c
  @{}}
    \toprule
    & \multicolumn{6}{c}{\textbf{Endpoint Method}} &
    & \multicolumn{6}{c}{\textbf{Request Parameter}} &
    & \multicolumn{6}{c}{\textbf{Response}} &
    & \multicolumn{6}{c}{\textbf{Parameter Constraint}}
    \\
    \cmidrule(lr){2-7}
    \cmidrule(lr){9-14}
    \cmidrule(lr){16-21}
    \cmidrule(lr){23-28}
    \textbf{REST API}
    & \textbf{Pr} & \textbf{BL} & \textbf{Re} & \textbf{BL} & \textbf{F1} & \textbf{BL} &
    & \textbf{Pr} & \textbf{BL} & \textbf{Re} & \textbf{BL} & \textbf{F1} & \textbf{BL} &
    & \textbf{Pr} & \textbf{BL} & \textbf{Re} & \textbf{BL} & \textbf{F1} & \textbf{BL} &
    & \textbf{Pr} & \textbf{BL} & \textbf{Re} & \textbf{BL} & \textbf{F1} & \textbf{BL}
    \\
    \midrule
    RealWorld / Ac           & 100 & 15.6 & 100 & 27.8 & 100 & 20.0 & & 82.1 & 26.9 & 100 & 30.4 & 90.2 & 28.6 & & 94.4 & 0.0 & 94.4 & 0.0 & 94.4 & N/A & & 100 & 11.5 & 94.4 & 16.7 & 97.1 & 13.6 \\
    Lin-CMS / Ac             & 100 & 27.3 & 100 & 9.4 & 100 & 14.0 & & 100 & 75.0 & 100 & 4.9 & 100 & 9.2 & & 100 & 0.0 & 100 & 0.0 & 100 & N/A & & 93.8 & 0.0 & 96.8 & 0.0 & 95.3 & N/A \\
    Management / Ac          & 100 & 42.1 & 100 & 13.6 & 100 & 20.5 & & 100 & 35.0 & 100 & 2.7 & 100 & 5.1 & & 100 & 0.0 & 100 & 0.0 & 100 & N/A & & 96.6 & 0.0 & 100 & 0.0 & 98.3 & N/A \\
    CMS / Ac                 & 100 & 16.7 & 100 & 8.8 & 100 & 11.5 & & 100 & 40.0 & 97.3 & 2.7 & 98.6 & 5.0 & & 97.1 & 0.0 & 97.1 & 0.0 & 97.1 & N/A & & 87.8 & 0.0 & 93.5 & 0.0 & 90.5 & N/A \\
    Accountill / Ex          & 100 & 83.3 & 100 & 100 & 100 & 90.9 & & 100 & 100 & 97.6 & 15.3 & 98.8 & 26.5 & & 95.0 & 0.0 & 95.0 & 0.0 & 95.0 & N/A & & 83.3 & 100 & 90.9 & 36.4 & 87.0 & 53.3 \\
    ProShop / Ex             & 100 & 35.7 & 100 & 20.8 & 100 & 26.3 & & 100 & N/A & 100 & 0.0 & 100 & N/A & & 100 & 0.0 & 100 & 0.0 & 100 & N/A & & 100 & N/A & 75.8 & 0.0 & 86.2 & N/A \\
    Proxy-Print / Rs         & 100 & 98.6 & 100 & 100 & 100 & 99.3 & & 94.2 & 95.8 & 97.0 & 82.1 & 95.6 & 88.5 & & 98.6 & 6.8 & 98.6 & 6.9 & 98.6 & 6.9 & & 93.5 & 100 & 91.8 & 41.8 & 92.7 & 59.0 \\
    Ur-Codebin / Rs          & 100 & 100 & 85.7 & 85.7 & 92.3 & 92.3 & & 100 & 42.9 & 87.5 & 37.5 & 93.3 & 40.0 & & 100 & 83.3 & 85.7 & 71.4 & 92.3 & 76.9 & & 100 & 100 & 91.7 & 41.7 & 95.7 & 58.8 \\
    Cassandra / Rs           & 96.2 & 100 & 100 & 100 & 98.0 & 100 & & 100 & 43.6 & 100 & 43.2 & 100 & 43.4 & & 96.5 & 17.3 & 100 & 16.4 & 98.2 & 16.8 & & 83.1 & N/A & 100 & 0.0 & 90.7 & N/A \\
    Education-1 / Rs         & 98.5 & 97.0 & 100 & 100 & 99.2 & 98.5 & & 97.0 & 89.1 & 100 & 84.5 & 98.5 & 86.8 & & 95.8 & 6.0 & 100 & 5.8 & 97.9 & 5.9 & & 88.0 & 100 & 93.6 & 41.5 & 90.7 & 58.6 \\
    Education-2 / \ding{55}  & 100 & \ding{55} & 100 & \ding{55} & 100 & \ding{55} & & 95.6 & \ding{55} & 99.5 & \ding{55} & 97.5 & \ding{55} & & 99.0 & \ding{55} & 99.0 & \ding{55} & 99.0 & \ding{55} & & 93.9 & \ding{55} & 92.5 & \ding{55} & 93.2 & \ding{55} \\
    Laravel-Blog / \ding{55} & 95.7 & \ding{55} & 95.7 & \ding{55} & 95.7 & \ding{55} & & 98.1 & \ding{55} & 89.5 & \ding{55} & 93.6 & \ding{55} & & 91.3 & \ding{55} & 91.3 & \ding{55} & 91.3 & \ding{55} & & 87.8 & \ding{55} & 92.3 & \ding{55} & 90.0 & \ding{55} \\
    \midrule
    \textbf{Average}         & 99.2 & 61.6 & 98.5 & 56.6 & 98.8 & 57.3 & & 97.2 & 60.9 & 97.4 & 30.3 & 97.2 & 37.0 & & 97.3 & 11.3 & 96.8 & 10.1 & 97.0 & 26.6 & & 92.3 & 51.4 & 92.8 & 17.8 & 92.3 & 48.7 \\
    \bottomrule
  \end{tabular}
\end{table*}

OOPS is implemented in Python, with approximately 3,000 lines of code for technology analysis, endpoint method extraction, and OAS generation. Compared to traditional OAS generation methods that require substantial manual implementation (e.g., Respector with over 10,000 lines of code), OOPS offers a more lightweight implementation. In the endpoint method extraction phase, we utilize Instructor \citep{liu2024instructor} to achieve structured output from LLMs. In the OAS generation phase, we employ JsonRef \citep{jsonref} to parse JSON references in LLM outputs, thereby enabling feedback and correction of reference errors; we use JSON Repair \citep{json-repair} to fix other JSON syntax errors in LLM outputs.

Regarding LLM selection, we comprehensively consider factors including the model's code capabilities, inference efficiency, API pricing, and function calling support. In the experiments for RQ1, RQ2, and RQ4, we selected the relatively balanced lightweight model GPT-5 mini. In the experiments for RQ3, we additionally tested Qwen3-Coder-Plus, GPT-5 nano, Gemini 2.0 Flash-Lite, and Gemini 3 Flash, with detailed information about these LLMs presented in Table~\ref{table-4-2}. We invoke the OpenAI-compatible APIs of these LLMs via the OpenAI SDK and set the temperature parameter to 0 to obtain deterministic results. All experiments were conducted on a server equipped with an Intel Xeon CPU, 128GB of memory, and the Ubuntu Server operating system.

\section{Experimental results}\label{section-5}

\subsection{Method effectiveness (RQ1)}\label{section-5-1}

\begin{table*}[ht]
  \scriptsize
  \centering
  \caption{Comparison of OOPS with 2 OOPS variants on 12 REST APIs. Since OOPS serially executes endpoint method extraction followed by OpenAPI specification generation, the Endpoint Method column compares OOPS with variant 1 (denoted as Vr1), while the Request Parameter, Response, and Parameter Constraint columns compare OOPS with variant 2 (denoted as Vr2).}\label{table-5-2}
  \vspace{0.2em}
  \begin{tabular}{@{} l @{\hskip 8pt}
    c @{\hskip 4pt} c @{\hskip 8pt} c @{\hskip 4pt} c @{\hskip 8pt} c @{\hskip 4pt} c @{} p{16pt} @{}
    c @{\hskip 4pt} c @{\hskip 8pt} c @{\hskip 4pt} c @{\hskip 8pt} c @{\hskip 4pt} c @{} p{16pt} @{}
    c @{\hskip 4pt} c @{\hskip 8pt} c @{\hskip 4pt} c @{\hskip 8pt} c @{\hskip 4pt} c @{} p{16pt} @{}
    c @{\hskip 4pt} c @{\hskip 8pt} c @{\hskip 4pt} c @{\hskip 8pt} c @{\hskip 4pt} c
  @{}}
    \toprule
    & \multicolumn{6}{c}{\textbf{Endpoint Method}} &
    & \multicolumn{6}{c}{\textbf{Request Parameter}} &
    & \multicolumn{6}{c}{\textbf{Response}} &
    & \multicolumn{6}{c}{\textbf{Parameter Constraint}}
    \\
    \cmidrule(lr){2-7}
    \cmidrule(lr){9-14}
    \cmidrule(lr){16-21}
    \cmidrule(lr){23-28}
    \textbf{REST API}
    & \textbf{Pr} & \textbf{Vr1} & \textbf{Re} & \textbf{Vr1} & \textbf{F1} & \textbf{Vr1} &
    & \textbf{Pr} & \textbf{Vr2} & \textbf{Re} & \textbf{Vr2} & \textbf{F1} & \textbf{Vr2} &
    & \textbf{Pr} & \textbf{Vr2} & \textbf{Re} & \textbf{Vr2} & \textbf{F1} & \textbf{Vr2} &
    & \textbf{Pr} & \textbf{Vr2} & \textbf{Re} & \textbf{Vr2} & \textbf{F1} & \textbf{Vr2}
    \\
    \midrule
    RealWorld        & 100 & 94.7 & 100 & 100 & 100 & 97.3 & & 82.1 & 95.7 & 100 & 95.7 & 90.2 & 95.7 & & 94.4 & 94.4 & 94.4 & 94.4 & 94.4 & 94.4 & & 100 & 94.1 & 94.4 & 88.9 & 97.1 & 91.4 \\
    Lin-CMS          & 100 & 100 & 100 & 18.8 & 100 & 31.6 & & 100 & 84.4 & 100 & 62.3 & 100 & 71.7 & & 100 & 18.8 & 100 & 18.8 & 100 & 18.8 & & 93.8 & 70.7 & 96.8 & 46.0 & 95.3 & 55.8 \\
    Management       & 100 & 100 & 100 & 23.7 & 100 & 38.4 & & 100 & 80.8 & 100 & 62.5 & 100 & 70.5 & & 100 & 27.1 & 100 & 27.1 & 100 & 27.1 & & 96.6 & 70.6 & 100 & 50.0 & 98.3 & 58.5 \\
    CMS              & 100 & 96.6 & 100 & 82.4 & 100 & 88.9 & & 100 & 85.4 & 97.3 & 46.7 & 98.6 & 60.3 & & 97.1 & 88.6 & 97.1 & 91.2 & 97.1 & 89.9 & & 87.8 & 76.5 & 93.5 & 56.5 & 90.5 & 65.0 \\
    Accountill       & 100 & 75.0 & 100 & 75.0 & 100 & 75.0 & & 100 & 80.9 & 97.6 & 44.7 & 98.8 & 57.6 & & 95.0 & 90.5 & 95.0 & 95.0 & 95.0 & 92.7 & & 83.3 & 76.0 & 90.9 & 86.4 & 87.0 & 80.9 \\
    ProShop          & 100 & 100 & 100 & 95.8 & 100 & 97.9 & & 100 & 100 & 100 & 100 & 100 & 100 & & 100 & 100 & 100 & 100 & 100 & 100 & & 100 & 95.5 & 75.8 & 63.6 & 86.2 & 76.4 \\
    Proxy-Print      & 100 & 23.1 & 100 & 12.5 & 100 & 16.2 & & 94.2 & 87.2 & 97.0 & 73.2 & 95.6 & 79.6 & & 98.6 & 94.5 & 98.6 & 95.8 & 98.6 & 95.2 & & 93.5 & 86.2 & 91.8 & 68.2 & 92.7 & 76.1 \\
    Ur-Codebin       & 100 & 100 & 85.7 & 85.7 & 92.3 & 92.3 & & 100 & 93.3 & 87.5 & 87.5 & 93.3 & 90.3 & & 100 & 42.9 & 85.7 & 42.9 & 92.3 & 42.9 & & 100 & 85.7 & 91.7 & 50.0 & 95.7 & 63.2 \\
    Cassandra        & 96.2 & 100 & 100 & 14.0 & 98.0 & 24.6 & & 100 & 88.4 & 100 & 80.0 & 100 & 84.0 & & 96.5 & 87.7 & 100 & 90.9 & 98.2 & 89.3 & & 83.1 & 71.1 & 100 & 65.3 & 90.7 & 68.1 \\
    Education-1      & 98.5 & 100 & 100 & 16.9 & 99.2 & 28.9 & & 97.0 & 88.5 & 100 & 79.4 & 98.5 & 83.7 & & 95.8 & 93.0 & 100 & 95.7 & 97.9 & 94.3 & & 88.0 & 81.7 & 93.6 & 52.1 & 90.7 & 63.6 \\
    Education-2      & 100 & 93.0 & 100 & 40.4 & 100 & 56.3 & & 95.6 & 84.5 & 99.5 & 82.8 & 97.5 & 83.7 & & 99.0 & 68.7 & 99.0 & 66.7 & 99.0 & 67.7 & & 93.9 & 93.6 & 92.5 & 65.7 & 93.2 & 77.2 \\
    Laravel-Blog     & 95.7 & 100 & 95.7 & 91.3 & 95.7 & 95.5 & & 98.1 & 45.7 & 89.5 & 84.2 & 93.6 & 59.3 & & 91.3 & 40.4 & 91.3 & 82.6 & 91.3 & 54.3 & & 87.8 & 39.6 & 92.3 & 92.3 & 90.0 & 55.4 \\
    \midrule
    \textbf{Average} & 99.2 & 90.2 & 98.5 & 54.7 & 98.8 & 61.9 & & 97.2 & 84.6 & 97.4 & 74.9 & 97.2 & 78.0 & & 97.3 & 70.5 & 96.8 & 75.1 & 97.0 & 72.2 & & 92.3 & 78.4 & 92.8 & 65.4 & 92.3 & 69.3 \\
    \bottomrule
  \end{tabular}
\end{table*}

\textbf{Endpoint Method:} As shown in Table~\ref{table-5-1}, OOPS achieves an average precision and recall of 99.2\% and 98.5\%. OOPS exhibits only a small number of false negatives in 2 REST APIs, specifically the login endpoints of Ur-Codebin and Laravel-Blog. The false negative in Ur-Codebin stems from its use of setFilterProcessesUrl to dynamically register endpoints in JWTAuthenticationFilter, rather than adopting the standard implementation approach with \texttt{@RequestMapping}; the false negative in Laravel-Blog arises from the use of the \texttt{authenticate} action when registering the login endpoint with \texttt{Route::post}, which the LLM failed to identify successfully. OOPS exhibits only a small number of false positives in 3 REST APIs, namely Cassandra, Laravel-Blog, and Education-1. In Cassandra, the LLM incorrectly identified a path prefix of an endpoint, resulting in the simultaneous output of both the correct version with the prefix and an incorrect version without the prefix; in Education-1, the LLM misidentified a static resource endpoint as a REST API endpoint; in Laravel-Blog, the LLM failed to completely exclude traditional HTTP endpoints that return HTML pages. In comparison, Respector performs well in endpoint method inference but similarly fails to identify the login endpoint in Ur-Codebin and suffers from false positives caused by static resources. ExpressO achieves high recall for Accountill but reports false positives for multiple non-REST API endpoints, including static resources, and exhibits numerous false negatives on ProShop. APICARV, due to its limited crawler coverage of endpoint methods, shows significantly lower precision and recall compared to OOPS.

\textbf{Request Parameter:} OOPS achieves an average precision and recall of 97.2\% and 97.4\%. OOPS exhibits recall below 97\% only in Ur-Codebin and Laravel-Blog, with the core reason being the failure to identify the corresponding endpoint methods. OOPS exhibits a relatively high number of false positives only in RealWorld. Through in-depth analysis, we found that the LLM has limitations in handling nested structures. For example, in the login functionality, the server uses the \texttt{LoginValidator} structure to parse the request body, which contains a \texttt{User} structure with 2 fields: \texttt{Email} and \texttt{Password}. The correct request body should contain a \texttt{user} field, under which the \texttt{email} and \texttt{password} fields are nested. However, the LLM incorrectly identified these 3 fields as parallel relationships, resulting in multiple false positives. Additionally, in Proxy-Print, the login endpoint passes username and password through query parameters, but the LLM incorrectly identified these parameters as request body parameters based on common patterns. It is worth noting that OOPS can effectively handle class inheritance relationships. For example, in the password modification functionality, OOPS can identify the \texttt{old\_password} field from \texttt{ChangePasswordSchema} and identify the \texttt{new\_password} and \texttt{confirm\_password} fields from its parent class \texttt{ResetPasswordSchema}, demonstrating its ability to understand complex code structures. In comparison, although Respector supports the inference of path parameters, query parameters, and request body parameters, it performs poorly in inferring parameter types, resulting in lower precision and recall than OOPS. ExpressO and APICARV lack the capability to infer request body parameters, leading to significantly decreased precision and recall.

\textbf{Response:} OOPS achieves an average precision and recall of 97.3\% and 96.8\%. OOPS exhibits recall below 94\% only in Ur-Codebin and Laravel-Blog, with the reason similarly being the failure to identify the corresponding endpoint methods. OOPS's errors mainly manifested in the following 2 aspects: (1) Inferring response status codes based on experience. For example, for endpoint methods that create resources or have no response content, although development frameworks default to returning a 200 status code, the LLM infers 201 and 204 status codes, based on RESTful best practices. (2) Inferring response types based on experience. For example, in some REST APIs, certain non-query endpoint methods return string-type messages, but the LLM incorrectly infers the response as an object type with a \texttt{message} field, causing a decrease in precision and recall. It is worth noting that OOPS can correctly infer fields that are not explicitly encoded. For example, in ProShop, OOPS can identify the \texttt{timestamps} configuration in Mongoose model definitions and automatically add the \texttt{createdAt} and \texttt{updatedAt} fields, demonstrating its ability to understand implicit framework behaviors. In comparison, unlike OOPS, Respector always infers responses as string types, resulting in poor precision and recall. The reason is that the corresponding REST API implementations serialize objects into strings before writing them to the response body, which manifests as string types at the development framework level and thus misleads Respector. However, for clients, the response is actually of object type. Since OOPS leverages LLMs to understand the functionality of endpoint methods at the semantic level, it can avoid the above problems. ExpressO and APICARV lack the capability to infer response bodies.

\textbf{Parameter Constraint:} OOPS achieves an average precision and recall of 92.3\% and 92.8\%. OOPS's recall shows a certain correlation with the choice of development framework. When REST APIs adopt Pydantic-based Schema definitions (such as Lin-CMS and Management), the LLM performs well in inferring parameter constraints, with recall reaching over 96\%. However, when REST APIs combine Mongoose to define data models (such as ProShop, Accountill, and CMS), the LLM is prone to false negatives, although the recall still outperforms baseline methods. OOPS may attach additional constraints based on parameter semantics that are not explicitly defined in the code, resulting in decreased precision. For example, for the \texttt{limit} parameter, OOPS attaches a constraint with a minimum value of 0, which, although beneficial for guiding correct client invocation, is not actually enforced by the server. It is worth noting that OOPS has limitations in extracting complex regular expressions. For example, in Lin-CMS, the \texttt{datetime} field involves a complex regular expression exceeding 150 characters in length. Although the LLM correctly identified the existence of this constraint, it made bracket-matching errors when performing the extraction, resulting in an invalid regular expression. In comparison, although Respector achieves 100\% precision, it exhibits numerous false negatives. ExpressO and APICARV lack the capability to infer parameter constraints, with correctly identified constraints only originating from default constraints of path parameters.

\begin{tcolorbox}

  \textbf{Summary for RQ1:} OOPS can effectively generate OAS for REST API implementations using different programming languages and development frameworks, outperforming baseline methods in correctly inferring endpoint method, request parameter, response, and parameter constraint, with average F1-scores improving by 41.5\%, 60.2\%, 70.4\%, and 43.6\%.

\end{tcolorbox}

\subsection{Ablation study (RQ2)}\label{section-5-2}

\begin{table*}[ht]
  \scriptsize
  \centering
  \caption{Comparison of OOPS performance across 5 different LLMs. Qwen3-Coder-Plus is denoted as qwen, GPT-5 mini as mini, GPT-5 nano as nano, Gemini 2.0 Flash-Lite as lite, and Gemini 3 Flash as flash. Each cell indicates the F1-score for OOPS using the corresponding LLM.}\label{table-5-3}
  \vspace{0.2em}
  \begin{tabular}{@{} l @{\hskip 14pt}
    c @{\hskip 7pt} c @{\hskip 7pt} c @{\hskip 7pt} c @{\hskip 7pt} c @{} p{14pt} @{}
    c @{\hskip 7pt} c @{\hskip 7pt} c @{\hskip 7pt} c @{\hskip 7pt} c @{} p{14pt} @{}
    c @{\hskip 7pt} c @{\hskip 7pt} c @{\hskip 7pt} c @{\hskip 7pt} c @{} p{14pt} @{}
    c @{\hskip 7pt} c @{\hskip 7pt} c @{\hskip 7pt} c @{\hskip 7pt} c
  @{}}
    \toprule
    & \multicolumn{5}{c}{\textbf{Endpoint Method}} &
    & \multicolumn{5}{c}{\textbf{Request Parameter}} &
    & \multicolumn{5}{c}{\textbf{Response}} &
    & \multicolumn{5}{c}{\textbf{Parameter Constraint}}
    \\
    \cmidrule(lr){2-6}
    \cmidrule(lr){8-12}
    \cmidrule(lr){14-18}
    \cmidrule(lr){20-24}
    \textbf{REST API}
    & \textbf{qwen} & \textbf{mini} & \textbf{nano} & \textbf{lite} & \textbf{flash} &
    & \textbf{qwen} & \textbf{mini} & \textbf{nano} & \textbf{lite} & \textbf{flash} &
    & \textbf{qwen} & \textbf{mini} & \textbf{nano} & \textbf{lite} & \textbf{flash} &
    & \textbf{qwen} & \textbf{mini} & \textbf{nano} & \textbf{lite} & \textbf{flash}
    \\
    \midrule
    RealWorld        & 100 & 100 & 76.9 & N/A & 100 & & 95.5 & 90.2 & 69.2 & N/A & 100 & & 94.4 & 94.4 & 71.8 & N/A & 94.4 & & 83.9 & 97.1 & 52.4 & N/A & 94.1 \\
    Lin-CMS          & 100 & 100 & 90.0 & 50.8 & 100 & & 96.8 & 100 & 83.2 & 29.5 & 100 & & 96.9 & 100 & 66.7 & 47.5 & 100 & & 92.6 & 95.3 & 73.7 & 22.5 & 95.9 \\
    Management       & 100 & 100 & 87.5 & 48.4 & 100 & & 95.2 & 100 & 81.2 & 22.5 & 100 & & 96.6 & 100 & 64.3 & 43.5 & 100 & & 90.8 & 98.3 & 71.3 & 18.9 & 98.6 \\
    CMS              & 100 & 100 & 90.7 & 81.0 & 100 & & 90.6 & 98.6 & 76.6 & 48.5 & 98.7 & & 97.1 & 97.1 & 88.0 & 81.0 & 100 & & 90.3 & 90.5 & 79.6 & 56.5 & 95.5 \\
    Accountill       & 100 & 100 & 83.3 & 78.4 & 100 & & 90.4 & 98.8 & 75.1 & 46.2 & 99.4 & & 100 & 95.0 & 77.6 & 75.5 & 100 & & 86.4 & 87.0 & 73.5 & 56.6 & 92.7 \\
    ProShop          & 100 & 100 & 98.0 & 92.3 & 100 & & 84.1 & 100 & 98.8 & 75.0 & 100 & & 87.5 & 100 & 98.0 & 90.6 & 100 & & 77.8 & 86.2 & 80.7 & 62.7 & 86.2 \\
    Proxy-Print      & 100 & 100 & 99.3 & 91.3 & 100 & & 92.1 & 95.6 & 91.0 & 52.0 & 93.9 & & 88.9 & 98.6 & 85.3 & 85.5 & 100 & & 90.7 & 92.7 & 82.3 & 51.7 & 64.3 \\
    Ur-Codebin       & 85.7 & 92.3 & 92.3 & 61.5 & 92.3 & & 93.3 & 93.3 & 81.5 & 46.2 & 93.3 & & 85.7 & 92.3 & 92.3 & 61.5 & 92.3 & & 95.7 & 95.7 & 47.1 & 37.5 & 85.7 \\
    Cassandra        & 98.0 & 98.0 & 98.0 & 97.0 & 100 & & 97.3 & 100 & 94.8 & 51.0 & 98.9 & & 94.6 & 98.2 & 98.2 & 97.2 & 100 & & 92.5 & 90.7 & 78.8 & 7.5 & 90.7 \\
    Education-1      & 99.2 & 99.2 & 97.7 & 93.8 & 99.2 & & 96.8 & 98.5 & 93.5 & 54.5 & 98.0 & & 95.0 & 97.9 & 90.5 & 91.0 & 99.3 & & 92.9 & 90.7 & 81.1 & 46.3 & 88.7 \\
    Education-2      & 100 & 100 & 72.5 & N/A & 100 & & 98.5 & 97.5 & 69.7 & N/A & 99.5 & & 100 & 99.0 & 73.5 & N/A & 100 & & 90.3 & 93.2 & 53.3 & N/A & 93.8 \\
    Laravel-Blog     & 77.2 & 95.7 & 24.6 & 42.6 & 100 & & 68.3 & 93.6 & 28.4 & 37.5 & 100 & & 13.3 & 91.3 & 24.1 & 36.9 & 87.0 & & 57.8 & 90.0 & 17.1 & 25.7 & 100 \\
    \midrule
    \textbf{Average} & 96.7 & 98.8 & 84.2 & 73.7 & 99.3 & & 91.6 & 97.2 & 78.6 & 46.3 & 98.5 & & 87.5 & 97.0 & 77.5 & 71.0 & 97.8 & & 86.8 & 92.3 & 65.9 & 38.6 & 90.5 \\
    \bottomrule
  \end{tabular}
\end{table*}

\textbf{Endpoint Method Extraction:} To evaluate the contribution of this step, we constructed Variant 1 of OOPS, which replaces the multi-agent workflow based on the API dependency graph with a simple ReAct-style agent. This agent is equipped with the same tool functions as OOPS (list directory, read file, search file) and directly extracts the list of REST API endpoint methods from the code repository. As shown in Table~\ref{table-5-2}, Variant 1 exhibits a 9.0\% decrease in precision, primarily due to its failure to correctly infer path prefixes of endpoint methods; the recall drops by 43.8\%, mainly because of limitations of the exploration strategy, i.e., the agent may terminate exploration after analyzing only a subset of code files and summarize endpoint methods based on incomplete information. Specifically, in REST API implementations where endpoint method definitions are concentrated in a small number of files (e.g., RealWorld, ProShop, Ur-Codebin, and Laravel-Blog), Variant 1 performs relatively well but still falls short of OOPS; whereas in REST API implementations where endpoint method definitions are scattered across multiple files, Variant 1 experiences a significant decline in recall. Since the performance degradation of Variant 1 in the endpoint method extraction step directly affects subsequent processes, leading to substantially lower quality in OpenAPI specification generation, Table~\ref{table-5-2} does not present its performance on request parameter, response, and parameter constraint inference.

\textbf{OpenAPI Specification Generation:} To evaluate the contribution of this step, we constructed Variant 2 of OOPS, which replaces the multi-agent workflow based on multi-stage generation and self-refine strategy with a single-pass generation approach, i.e., directly calling an LLM to generate Operation objects compliant with the OpenAPI 3.0.4 specification. Since Variant 2 employs the same endpoint method extraction step as OOPS, both achieve identical performance in endpoint method inference. Therefore, Table~\ref{table-5-2} does not present this part of the performance. Experimental results demonstrate that Variant 2 exhibits performance degradation across request parameter, response, and parameter constraint inference. This degradation can be attributed to 2 factors: (1) some Operation objects output by the LLM are discarded due to non-compliance with the OAS definition; (2) the single-pass generation strategy requires the LLM to simultaneously perform request parameter structure analysis and response structure analysis, which is a highly complex task, resulting in outputs that, even when syntactically correct, are of significantly lower semantic quality than those produced by the multi-stage generation approach. Notably, in the request parameter inference task for RealWorld, Variant 2 achieves higher precision than OOPS, likely because RealWorld's response format follows a nested structure of \texttt{\{"user":\{"id":"","username":""\}\}}, which is represented in the code as \texttt{gin.H\{"user":...\}}, a pattern that is highly recognizable to the LLM. Once the LLM correctly infers the response structure, it can leverage this structural information to infer that the request body has a similar structure.

\begin{tcolorbox}

  \textbf{Summary for RQ2:} Both key steps contribute significantly to OOPS's performance. Designs in the endpoint method extraction step achieve a 36.9\% average F1-score improvement. Designs in the OpenAPI specification generation step achieve average F1-score improvements of 19.2\%, 24.8\%, and 23.0\% for request parameter, response, and parameter constraint inference.

\end{tcolorbox}

\subsection{Impact of different LLM choices (RQ3)}\label{section-5-3}

\begin{table*}[ht]
  \scriptsize
  \centering
  \caption{Overhead analysis of OOPS. Calls represents the number of LLM invocations, Avg In and Max In denote the average and maximum input tokens to LLMs, Avg Out and Max Out denote the average and maximum output tokens from LLMs, Max Dep indicates the maximum number of file associations during endpoint method extraction, and Cost represents the total cost of generating OAS using OOPS.}\label{table-5-4} \vspace{0.2em}
  \begin{tabular}{@{} l @{\hskip 18pt}
    c @{\hskip 12pt} c @{\hskip 12pt} c @{\hskip 12pt} c @{\hskip 12pt} c @{} p{18pt} @{}
    c @{\hskip 12pt} c @{\hskip 12pt} c @{\hskip 12pt} c @{\hskip 12pt} c @{} p{18pt} @{}
    c @{\hskip 12pt} c @{\hskip 12pt}
  @{}}
    \toprule
    & \multicolumn{5}{c}{\textbf{Endpoint Method Extraction}} &
& \multicolumn{5}{c}{\textbf{OpenAPI Specification Generation}} &
    \\
    \cmidrule(lr){2-6}
    \cmidrule(lr){8-12}
    \textbf{REST API}
    & \textbf{Calls} & \textbf{Avg In} & \textbf{Max In} & \textbf{Avg Out} & \textbf{Max Out} &
    & \textbf{Calls} & \textbf{Avg In} & \textbf{Max In} & \textbf{Avg Out} & \textbf{Max Out} &
    & \textbf{Max Dep} & \textbf{Cost}
    \\
    \midrule
    RealWorld        & 96 & 1.15K & 3.42K & 0.66K & 3.49K & & 212 & 3.78K & 8.83K & 0.40K & 2.69K & & 2 & \$0.23 \\
    Lin-CMS          & 259 & 1.02K & 4.97K & 0.68K & 3.77K & & 388 & 4.26K & 12.01K & 0.34K & 2.59K & & 3 & \$0.48 \\
    Management       & 329 & 1.15K & 4.40K & 0.72K & 4.12K & & 714 & 4.53K & 13.27K & 0.45K & 3.46K & & 3 & \$0.90 \\
    CMS              & 421 & 1.01K & 5.33K & 0.55K & 4.11K & & 246 & 2.50K & 5.82K & 0.76K & 3.17K & & 2 & \$0.26 \\
    Accountill       & 242 & 0.95K & 3.46K & 0.56K & 4.24K & & 92 & 1.46K & 3.52K & 0.75K & 2.54K & & 3 & \$0.09 \\
    ProShop          & 331 & 0.99K & 2.75K & 0.57K & 4.09K & & 100 & 2.00K & 3.37K & 0.77K & 2.66K & & 3 & \$0.13 \\
    Proxy-Print      & 466 & 1.58K & 11.94K & 0.75K & 4.48K & & 300 & 5.59K & 16.13K & 0.74K & 3.10K & & 1 & \$0.60 \\
    Ur-Codebin       & 93 & 0.63K & 2.48K & 0.69K & 4.46K & & 87 & 2.68K & 7.01K & 0.35K & 2.59K & & 1 & \$0.07 \\
    Cassandra        & 348 & 1.92K & 9.20K & 0.72K & 5.06K & & 161 & 4.42K & 12.44K & 0.83K & 3.22K & & 1 & \$0.34 \\
    Education-1      & 885 & 1.79K & 10.93K & 0.75K & 4.73K & & 253 & 5.15K & 15.32K & 0.78K & 3.27K & & 1 & \$0.72 \\
    Education-2      & 336 & 1.04K & 6.73K & 0.64K & 4.70K & & 772 & 3.31K & 9.22K & 0.49K & 3.67K & & 3 & \$0.73 \\
    Laravel-Blog     & 697 & 0.67K & 3.86K & 0.59K & 7.63K & & 488 & 2.17K & 8.78K & 0.47K & 3.44K & & 2 & \$0.38 \\
    \midrule
    \textbf{Maximum} & 885 & 1.92K & 11.94K & 0.75K & 7.63K & & 772 & 5.59K & 16.13K & 0.83K & 3.67K & & 3 & \$0.90 \\
    \textbf{Minimum} & 93 & 0.63K & 2.48K & 0.55K & 3.49K & & 87 & 1.46K & 3.37K & 0.34K & 2.54K & & 1 & \$0.07 \\
    \textbf{Average} & 375.25 & 1.16K & 5.79K & 0.66K & 4.57K & & 317.75 & 3.49K & 9.64K & 0.59K & 3.03K & & 2.08 & \$0.41 \\
    \bottomrule
  \end{tabular}
\end{table*}

\textbf{Endpoint Method:} As shown in Table~\ref{table-5-3}, Gemini 3 Flash outperforms GPT-5 mini in inferring endpoint methods with a higher F1-score, primarily due to lower false positive rates on Cassandra and Laravel-Blog. Qwen3-Coder-Plus achieves F1-scores exceeding 96\% on most REST APIs, but its overall performance is slightly lower than GPT-5 mini due to multiple false positives in Ur-Codebin and Laravel-Blog. In contrast, the more lightweight GPT-5 nano and Gemini 2.0 Flash-Lite exhibit poor performance in inferring endpoint methods. Notably, Gemini 2.0 Flash-Lite failed to correctly complete file dependency analysis when analyzing RealWorld and Education-2, resulting in the failure to construct API dependency graphs and thus yielding entirely incorrect endpoint method inferences.

\textbf{Request Parameter:} Gemini 3 Flash outperforms GPT-5 mini in inferring request parameters with a higher F1-score, especially for RealWorld, where Gemini 3 Flash accurately identified the nested structure of request bodies, reducing the false positive rate. Qwen3-Coder-Plus also outperforms GPT-5 mini in inferring request parameters for RealWorld, but its overall performance is lower than GPT-5 mini. This is attributed to numerous false positives in Laravel-Blog as well as poor inference performance on three REST API implementations based on Mongoose-defined data models: ProShop, Accountill, and CMS. In contrast, the more lightweight GPT-5 nano and Gemini 2.0 Flash-Lite show significant performance degradation.

\textbf{Response:} Gemini 3 Flash also outperforms GPT-5 mini in inferring responses with a higher F1-score, primarily due to higher accuracy in type inference. Qwen3-Coder-Plus exhibits overall performance lower than GPT-5 mini in inferring responses, particularly on Laravel-Blog, where its F1-score is only 13.3\%. This is attributed to both numerous false positives during endpoint method inference and the failure to understand Laravel's resource-based response mechanism for REST API implementation. Additionally, compared to inferring request parameters, Gemini 2.0 Flash-Lite performs better in inferring responses, while other LLMs show minimal performance differences between these 2 dimensions.

\textbf{Parameter Constraint:} GPT-5 mini achieves the best F1-score in inferring parameter constraints, rather than Gemini 3 Flash, which achieves the best performance in other evaluation dimensions. This is primarily because Gemini 3 Flash performs poorly when handling Proxy-Print and Ur-Codebin based on Spring Boot. Specifically, Gemini 3 Flash may be misled by no-argument constructors, leading to the loss of numerous constraints and consequently a decrease in F1-score. Qwen3-Coder-Plus may have adopted a more conservative strategy when inferring parameter constraints, resulting in lower performance than GPT-5 mini due to multiple false negatives. Due to limitations in request parameter inference capability, GPT-5 nano and Gemini 2.0 Flash-Lite also exhibit poor performance in parameter constraint inference.

\begin{tcolorbox}

  \textbf{Summary for RQ3:} The performance of OOPS is influenced by base LLMs. Gemini 3 Flash performs best in inferring endpoint methods, request parameters, and responses, while GPT-5 mini excels in parameter constraint inference. Qwen3-Coder-Plus shows potential but with lower overall performance, whereas the more lightweight GPT-5 nano and Gemini 2.0 Flash-Lite exhibit poor performance.

\end{tcolorbox}

\subsection{LLM Context Efficiency (RQ4)}\label{section-5-4}

\textbf{Endpoint Method Extraction:} As shown in Table~\ref{table-5-4}, this step requires an average of 375.25 LLM calls for each REST API, with the average input tokens not exceeding 1.92K and a maximum of 11.94K, observed in Proxy-Print, which is attributed to the large size of some of its files. Although some REST API implementations, such as Lin-CMS, Laravel-Blog, and Education-1, contain a large number of code files, through file filtering, OOPS can effectively reduce unnecessary file analysis, thereby maintaining a low average input length. The average output tokens in this step do not exceed 0.75K, with a maximum of 7.63K, observed in Laravel-Blog, which is attributed to some of its files containing a large number of API entries.

\textbf{OpenAPI Specification Generation:} As shown in Table~\ref{table-5-4}, this step requires an average of 317.75 LLM calls for each REST API, with the average input tokens not exceeding 5.59K and a maximum of 16.13K. Compared to the endpoint method extraction step, the average input tokens in this step increase significantly. This increase is primarily because when inferring request parameters and response structures, the files that the endpoint methods depend on need to be input to the LLM together; additionally, the LLM agent may further read other relevant files as needed. The average output tokens in this step do not exceed 0.83K, with a maximum of 3.67K, indicating that the multi-stage generation strategy can effectively control the output length of individual LLM calls.

\textbf{File Association Scale:} Associating files through the API dependency graph is a key design of OOPS to address the context length limitations of LLMs. In the endpoint method extraction step, the maximum number of file associations for all REST API implementations does not exceed 3, indicating that the API dependency graph constructed by OOPS has a shallow hierarchical structure, which can effectively avoid context overflow caused by excessive file associations. Specifically, for REST API implementations that adopt multi-level nested route definitions, such as RealWorld, ProShop, Accountill, and Lin-CMS, the file dependency depth is 2 or 3 levels; while for REST API implementations based on Spring Boot and Jersey that adopt annotation-driven development patterns, such as Proxy-Print, Ur-Codebin, Cassandra, and Education-1, the file dependency depth is only 1 level. Such shallow dependency relationships not only reduce context overhead but also improve the inference efficiency of LLMs.

\begin{tcolorbox}

  \textbf{Summary for RQ4:} The average total cost for OOPS to generate a complete OAS for each REST API implementation is \$0.41. Although a large number of LLM calls are required, OOPS effectively controls token consumption through optimization strategies, keeping the input and output tokens of individual calls far below the limits of modern LLMs.

\end{tcolorbox}

\section{Threats to validity}\label{section-6}

\subsection{Threats to construct validity}\label{section-6-1}

The LLMs we utilized are trained on a vast array of open-source datasets, which poses a potential risk of data leakage \citep{al2024traces}. In the context of our task, the training data of LLMs might include the open-source REST API implementations and the OAS authored by their developers, which could threaten the generalizability of our experimental findings. Despite this concern, we did not observe any instances where the outputs of OOPS were identical to the OAS authored by developers of the 8 open-source REST APIs. Moreover, our experimental dataset includes 4 private REST API implementations, whose OAS are not included in the LLMs' training data. The experimental results demonstrate that OOPS can still accurately generate the OAS for these REST API implementations. Therefore, we consider this threat to be limited to OAS generation tasks, especially compared to tasks such as code generation or code repair.

\subsection{Threats to internal validity}\label{section-6-2}

LLMs possess multiple hyperparameters, and variations in these settings may impact the experimental outcomes, posing a threat to internal validity. In particular, the output of LLMs is significantly influenced by the temperature parameter, which directly affects the randomness of the model's output. An excessively high temperature may lead to substantial variability in the generated text. To mitigate this threat, we set the temperature to 0 to enhance the consistency of the LLMs' output, thereby reducing the variance in results. Additionally, research on prompt engineering indicates that different prompting methods can significantly affect the quality of LLMs' outputs. To address this threat, we carefully designed the prompt templates used by OOPS at each step based on best practices and established design principles from both academia and industry, such as prompt formatting, specific descriptions, role-playing, chain of thought, and structured output control.

\subsection{Threats to external validity}\label{section-6-3}

The primary threat to external validity arises from the dataset used in our experiments. To support the technology-agnostic nature of our approach, the REST API implementations used to test OOPS involve various programming languages and development frameworks. However, these projects do not cover the entire spectrum of programming languages and development frameworks, and thus, the applicability of our results to all scenarios may be limited. Additionally, OOPS has only been tested on projects using publicly available technologies, and it remains uncertain whether OOPS can effectively operate on projects that employ proprietary development frameworks or less common programming languages.

\section{Related work}\label{section-7}

\subsection{Traditional OAS generation techniques}\label{section-7-1}

Currently, several OAS generation techniques based on annotations or comments exist, such as Gin-Swagger \citep{gin-swagger}, Springfox \citep{springfox}, Swagger-JSDoc \citep{swagger-jsdoc}, and Flasgger \citep{flasgger}. After developers provide specific annotations or comments, these techniques scan the code to produce OAS. Consequently, the quality of OAS depends heavily on developer effort. Another approach, represented by APICARV \citep{yandrapally2023carving}, involves executing UI tests or employing crawlers on REST APIs to collect HTTP logs containing API request and response information, then inferring path parameters and constructing OAS based on these logs. When UI tests are unavailable, APICARV utilizes Crawljax \citep{mesbah2012crawling} as its crawler. More advanced crawlers, such as the data-driven Black Widow \citep{eriksson2021black} or task-driven YuraScanner \citep{stafeev2024yurascanner}, can explore client states more extensively to gather valuable HTTP logs. However, these approaches rely on the existence of web clients and can only discover endpoints that are defined in the web client. In addition to approaches based on web clients, ExpressO \citep{serbout2022expresso}, based on server code, employs proxy injection combined with static analysis to generate OAS, requiring the REST API to be runnable. Recent techniques that do not involve LLMs generate OAS entirely through static analysis of server code \citep{huang2024generating,lercher2024generating}, thus eliminating the need to run REST APIs. These methods have also been applied to detect breaking changes in REST APIs \citep{lercher2025autoguard}, but all focus exclusively on the Java REST API frameworks Spring Boot and Jersey.

\subsection{LLM-based OAS generation techniques}\label{section-7-2}

The advent of LLMs introduces novel approaches to OAS generation. Chaplia \citep{chaplia2024extracting} employs LLM agents to analyze Node.js-based microservice implementations and utilizes traditional static analysis to determine inter-file references for OAS generation. LRASGen \citep{deng2025lrasgen} identifies key code files containing REST API definitions using human-expert-crafted regular expressions, then leverages an LLM workflow to generate OAS, with support for several programming languages and development frameworks. In contrast to these code-based methods, OASBuilder \citep{lazar2025generating} uses LLMs to generate OAS from online API documentation, offering complementary capabilities to the aforementioned approaches. Additionally, some studies leverage LLMs to generate formal program specifications \citep{ma2025specgen,wen2024enchanting}, exploiting similar code understanding and generation capabilities. Unlike OAS generation tasks, these specification generation tasks focus more on deeply analyzing individual function behaviors rather than understanding the externally exposed interfaces of entire projects.

\subsection{Repository-level code understanding}\label{section-7-3}

Analyzing individual functions or code files using LLMs alone is often insufficient for complex software engineering tasks. In repository-level code understanding, LLMs face significant challenges due to context length limitations \citep{zan2024codes,zhang2025hierarchical,liang2025can}. Existing research primarily addresses code generation \citep{phan2025repohyper,liu2024stall+,deng2024r2c2,zhang2025coderag,bui2024rambo} and code repair \citep{bairi2024codeplan,ma2025alibaba,liu2025codexgraph,ouyang2024repograph} tasks. Some approaches employ RAG techniques optimized for code files, enabling LLMs to perform cross-file analysis. Other studies utilize AST-based traditional static analysis methods to model dependency relationships as graph structures, thereby enabling LLMs to accurately capture global repository information. Similar to these works, our approach also employs graph structures to model dependencies. However, our work differs in modeling granularity and construction methodology: our API dependency graph adopts a coarse-grained design that maintains only essential information for identifying endpoint methods. Furthermore, we construct the API dependency graph entirely using LLM agent workflows, thereby achieving superior generalizability across different programming languages and development frameworks.

\subsection{Generalizable LLM4SE}\label{section-7-4}

Many LLM-driven software engineering tasks target specific programming languages. For instance, in test case generation, ChatTester \citep{yuan2024evaluating} supports only Java, TestPilot \citep{schafer2023empirical} is designed for JavaScript, and CityWalk \citep{zhang2025citywalk} primarily targets C++. In contrast, other LLM-based approaches emphasize generalizability across different technologies. Fuzz4All \citep{xia2024fuzz4all} employs LLMs as engines for input generation and mutation, enabling fuzzing for systems under test that accept formal languages as input. ExecutionAgent \citep{bouzenia2025you} integrates meta-prompts with feedback-driven iterative optimization, enabling LLM agents to build arbitrary projects from source code and execute their test suites. Inspired by these efforts, our work adopts a technology-agnostic design and demonstrates strong performance across different programming languages and development frameworks.

\section{Conclusion}\label{section-8}

In this paper, we propose OOPS, a novel approach that leverages LLM agent workflows to analyze the source code of REST APIs and generate OAS, supporting various server-side technologies without requiring human experts for technology-specific adaptations. By constructing an API dependency graph for file association, OOPS addresses the context length limitations of LLMs during endpoint method extraction. Through multi-stage generation and self-refine, OOPS alleviates various hallucinations of LLMs during OAS generation. To validate the effectiveness of OOPS, we evaluated it on 12 REST API implementations, encompassing open-source software and private projects, involving 5 programming languages and 8 development frameworks. The results demonstrate that OOPS achieves superior performance and generalizability compared to state-of-the-art OAS generation methods.

Since 4XX and 5XX status codes are equally valuable for guiding clients to correctly invoke REST APIs, especially as 4XX status codes reveal predictable client errors, in future work, we plan to extend OOPS's support for error status codes, particularly identifying error responses handled at the middleware layer, to provide more comprehensive OAS. Furthermore, to address semantic hallucinations caused by misunderstanding of business logic, we plan to draw on the idea of ExecutionAgent \citep{bouzenia2025you}, leveraging LLM agents to automatically deploy REST APIs and construct actual requests, thereby dynamically validating the correctness of the generated OAS and further mitigating semantic hallucinations of LLMs. Building upon these improvements, we will conduct empirical studies on a broader range of programming languages and development frameworks. Additionally, we aim to extend OOPS to generate functional summaries for each endpoint method and integrate it with LLM-based REST API testing methods such as LogiAgent \citep{zhang2025logiagent} and LlamaRestTest \citep{kim2025llamaresttest}, thereby enhancing the applicability and effectiveness of REST API testing.

\section*{CRediT authorship contribution statement}

Hao Chen: Conceptualization, Methodology, Software, Data curation, Writing - original draft, Writing - review \& editing, Validation. Yunchun Li: Resources, Supervision. Chen Chen: Writing - review \& editing. Fengxu Lin: Resources. Wei Li: Writing - review \& editing, Supervision.

\section*{Declaration of competing interest}

The authors declare that they have no known competing financial interests or personal relationships that could have appeared to influence the work reported in this paper.

\section*{Acknowledgments}

The authors express their gratitude to those who provided datasets and thank the editors and anonymous reviewers for their insightful comments and suggestions. This work was supported by the National Key Research and Development Program of China (Grant No. 2022YFB4501604).

\section*{Declaration of Generative AI and AI-assisted technologies in the writing process}

During the preparation of this work, the authors utilized ChatGPT and Claude to enhance language quality and readability. After using these tools, the authors reviewed and edited the content as needed and took full responsibility for the content of this publication.

\section*{Data availability}

The source code and the public datasets used in this study are available at \url{https://github.com/jsi-cst/OOPS}.

\bibliographystyle{elsarticle-harv}

\bibliography{refs}

\end{document}